\documentclass[10pt]{iopart}
\pdfoutput=1
\usepackage{amssymb}
\usepackage{graphicx}
\usepackage{cite,booktabs}
\begin{document}
\title[]{Configurational constraints on glass formation in the liquid calcium aluminate system}
\author{James W. E. Drewitt$^{1}$, Sandro Jahn$^{2}$, Louis Hennet$^{3}$}
\address{$^1$ School of Earth Sciences, University of Bristol, Wills Memorial Building, Queen's Road, Bristol, BS8 1RJ, United Kingdom}
\address{$^2$ Institute of Geology and Mineralogy, University of Cologne, Zuelpicher Str. 49b, 50674 Cologne, Germany}
\address{$^3$ Conditions Extr\^{e}mes et Mat\'{e}riaux : Haute Temp\'{e}rature et Irradiation, CEMHTI-CNRS,
Universit\'{e} d'Orl\'{e}ans, 1d avenue de la Recherche Scientifique, 45071 Orl\'{e}ans cedex 2, France}
\ead{james.drewitt@gmail.com}
\begin{abstract}
We report new time-resolved synchrotron x-ray diffraction (SXRD) measurements to track structural transformations in calcium-aluminate (CaO)$_{x}$(Al$_{2}$O$_{3}$)$_{1-x}$ liquids during glass formation, and review recent progress in neutron diffraction with isotope substitution (NDIS) experiments, combined with aspherical ion model molecular dynamics (AIM-MD) simulations, to identify the atomic-scale configurational constraints on glass-forming ability. The time-resolved measurements reveal substantial changes in ordering on short- and intermediate-range occurring during supercooling. In the equimolar composition $x=0.5$ (CA), the liquid undergoes a remarkable structural re-organisation on vitrification as over coordinated AlO$_{5}$ polyhedra and oxygen triclusters breakdown to form a network of predominantly corner-shared AlO$_{4}$ tetrahedra. This is accompanied by the formation of branched chains of edge- and face-sharing Ca-centred CaO$_{y}$ polyhedra contributing to cationic ordering on intermediate length-scales. The Ca-rich end-member of the glass-forming system $x=0.75$ (C3A) is largely composed of AlO$_{4}$ tetrahedra, but $\sim10$\,\% unconnected AlO$_{4}$ monomers and Al$_{2}$O$_{7}$ dimers are present, representing a threshold after which the glass can no longer support the formation of an infinitely connected network. Overall, the AIM-MD simulations are in excellent agreement with the SXRD and NDIS experiments suggesting an accurate potential model. However, small discrepancies between the simulated glass structures and experimental measurements are apparent indicating a small degree of liquid-like ordering persists in the simulated glass trajectories. This may be due to the short simulation time-scales which are unrepresentative of the viscous kinetic processes involved in supercooling and glass formation. One approach to improve future models could be the integration of rare event sampling techniques into MD simulation codes to massively extend equilibration time-scales and more accurately model vitrification and structural configurations in real glass systems.
\end{abstract}
\submitto{\JSTAT}
\maketitle
\ioptwocol
\section{Introduction}\label{Intro}
Aluminate glasses are fundamentally intriguing materials in which aluminium, a non-traditional network glass former, can assume an array of local structural chemistries \cite{McMillan96}. Aluminate glasses are also promising candidates as infra-red waveguide and sensor materials, and as hosts for optically active rare earth ions in photonics technology \cite{Shelby89,Weber98,Aizawa04,Haladejova16,Eeckhout10}.  Zachariasen's rules of glass formation outline four key atomic-scale characteristics of a glass former; 1) each oxygen atom is linked to no more than two glass-forming atoms, 2) the coordination number of the glass-forming atoms is small (three of four), 3) the polyhedra share corners, not edges or faces, and are 4) linked in a continuous 3-dimensional network \cite{Zachariasen32}.  Although often used as an additive in oxide glasses, pure alumina (Al$_{2}$O$_{3}$) itself does not form a glass as it is insufficiently oxygen-rich to form a continuous random network of corner-shared tetrahedra. To compensate for the oxygen deficiency in pure liquid Al$_{2}$O$_{3}$, oxygen atoms form clusters with three or more Al-units and one-third of all Al-polyhedra are coordinated by five or six oxygen atoms with considerable edge-sharing, features which break Zachariasen's rules \cite{Skinner13a}.  However, the introduction of other oxide components, such as CaO, increases the O:Al ratio which promotes the formation of higher fractions of AlO$_{4}$ tetrahedra to facilitate vitrification. 
\begin{figure}
\centering
\includegraphics[width=0.48\textwidth]{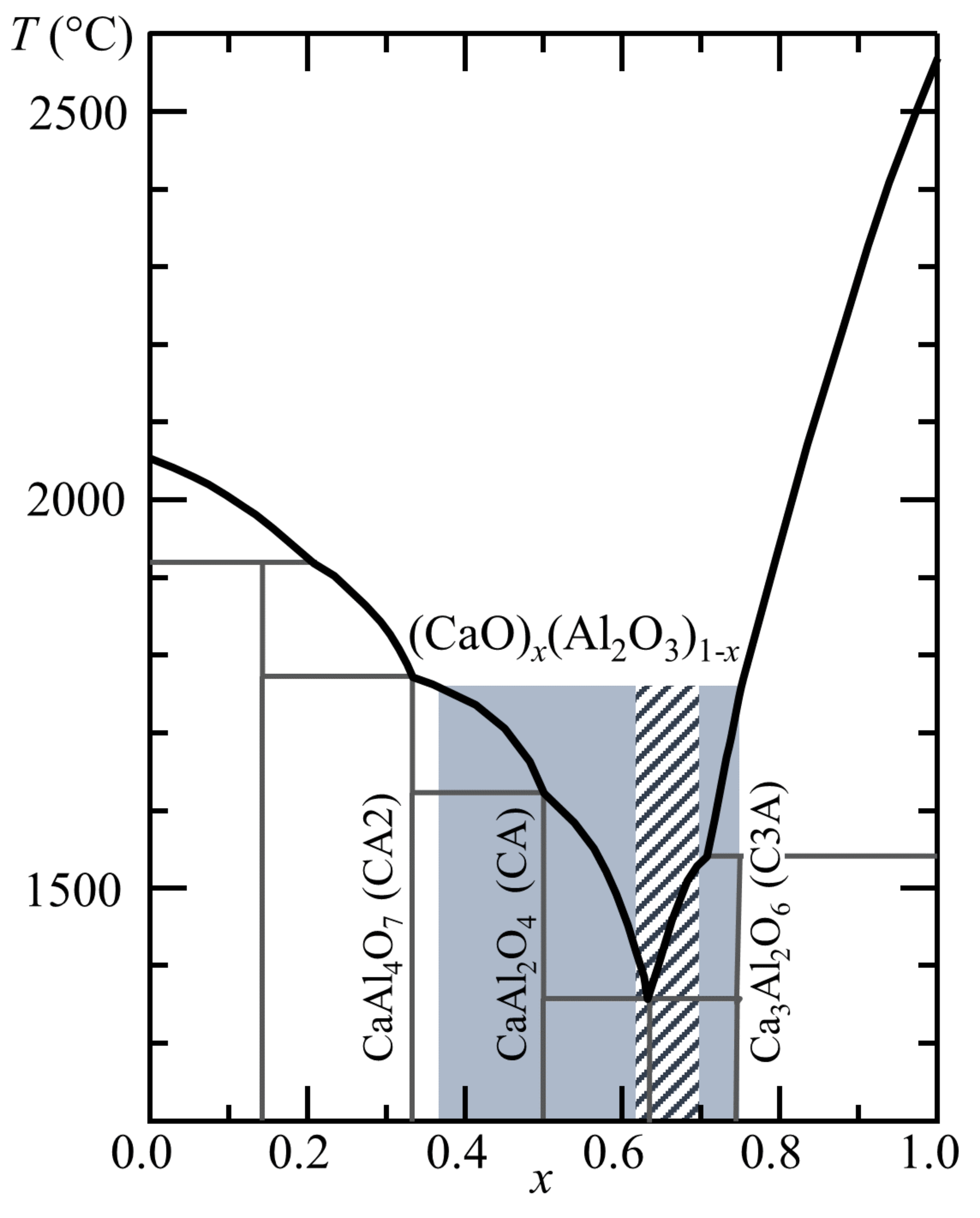}
\caption{Liquidus (black curve) \cite{Nurse65}, conventional glass-forming region (hatched area) \cite{Shelby89}, and extended glass forming region by containerless processing (solid filled area) \cite{Massiot98} for the system (CaO)$_{x}$(Al$_{2}$O$_{3}$)$_{1-x}$. Using rapid (splat) quenching it is possible to extend the glass-forming region further to $x=0.18$ \cite{McMillan83}. \label{Fig1-GFR}}
\end{figure}
Using conventional methods, (CaO)$_{x}$(Al$_{2}$O$_{3}$)$_{1-x}$ liquids can be vitrified to colourless transparent glasses over a relatively narrow region centred around $x=0.65$ close to the eutectic \cite{Shelby89}. Containerless processing enables the fabrication of glasses that cannot be formed using traditional methods by eliminating the possibility of chemical reactions between high temperature liquid oxides and containment materials and suppressing heterogeneous nucleation to promote deep supercooling. Using aerodynamic levitation with laser-heating, it is possible to extend this glass forming region to $0.37\lesssim x \lesssim 0.75$ \cite{Massiot98} (figure \ref{Fig1-GFR}). 

\begin{figure}
\centering
\includegraphics[width=0.48\textwidth]{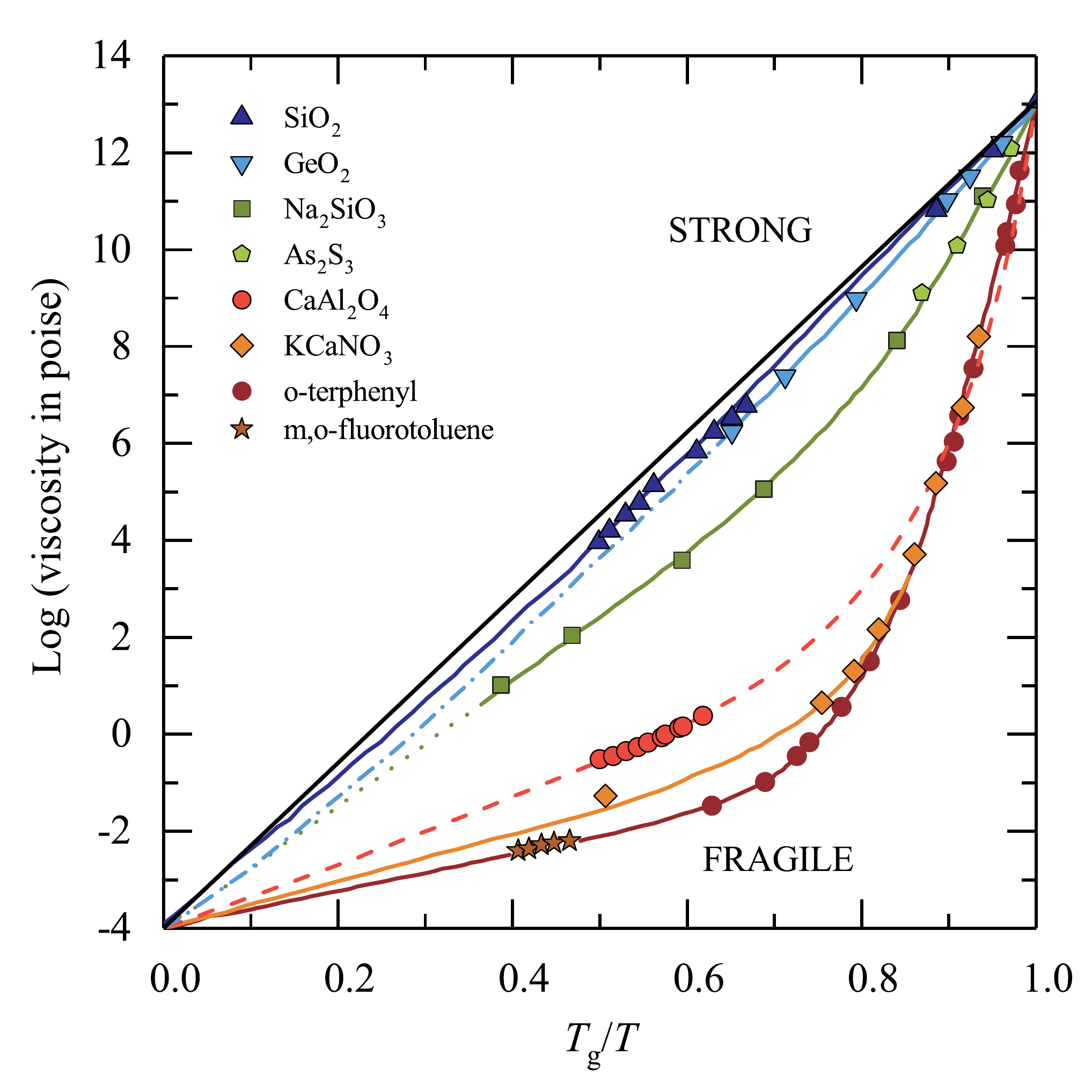}
\caption{Angell representation showing the viscosity-temperature dependence for selected glass-forming liquids (adapted from Angell (1995) \cite{Angell95}). The viscosities reported for liquid CaAl$_{2}$O$_{4}$ (CA) by Urbain (1983) \cite{Urbain83} are also shown. A Vogel-Tammann-Fulcher fit to the CA data, shown by the light red dashed curve, provides a fragility index $m=116$ \cite{Drewitt12b}, characteristic of a ``fragile'' liquid \cite{Bohmer93}.\label{Fig2-Angell}}
\end{figure}
Glass forming liquids may be classified in terms of kinetic fragility, a concept introduced by Angell \cite{Angell95}, in which ``strong'' liquids exhibit an approximately Arrhenius viscosity temperature dependence and ``fragile'' liquids exhibit non-Arrhenius behaviour characterised by a drastic slow-down in their dynamical properties as they approach the glass transition temperature $T_{\rm{g}}$.  A Vogel-Tammann-Fulcher \cite{Vogel21,Fulcher25,Tammann26} fit to the macroscopic viscosity data of liquid CaAl$_{2}$O$_{4}$ ($x=0.5$ CA) \cite{Urbain83} (see figure \ref{Fig2-Angell}) provides a fragility index, defined as the gradient at $T_{\rm{g}}$, of $m=116$ \cite{Drewitt12b} which is characteristic of a fragile liquid \cite{Bohmer93}.  By comparison, the canonical strong network glass-forming liquid SiO$_{2}$ has a fragility index $m=20$ \cite{Bohmer93}. The behaviour of fragile liquids arises from the larger range of densely packed potential energy minima in configurational space compared to strong liquids resulting from a higher degree of short-range disorder associated with local coordination and geometrical variation, and intermediate-range disorder associated with the connectivities between cation coordination polyhedra or extended channel structures \cite{Angel85,Debenedetti01,Greaves07}. In contrast, strong glass-forming liquids such as SiO$_{2}$ exhibit stable local coordination environments and self-reinforcing three-dimensional networks which restrict the number of available configurations and generate broad deep minima in the potential energy landscape. Mode coupling theory predicts a critical temperature $T_{\rm{m}}>T_{\rm{c}}>T_{\rm{g}}$, where $T_{\rm{m}}$ denotes the melting temperature and the critical temperature (designated the dynamical crossover temperature) is typically $T_{\rm{c}}\simeq1.2T_{\rm{g}}$ \cite{Gotze92,Anderson05,Ossi06}. Above $T_{\rm{c}}$, characteristically liquid diffusive motions dominate, whereas below $T_{\rm{c}}$ the transition towards a solid and dynamic arrest begins. On cooling through the dynamical crossover region the supercooled liquid encounters high potential energy barriers compared to thermal energies and is increasingly unable to explore the full-range of configurational states \cite{Goldstein69}. As a result, the system becomes trapped in a deep local energy minimum. It is therefore evident that a comprehensive understanding of liquid-state structure and its evolution during supercooling to form a solid glass is an important prerequisite for understanding the nature of glass-formation in fragile liquids.

The inherent structural disorder of liquids makes their atomic-scale structures difficult to characterise, although chemical bonding constraints can lead to a high degree of ordering on short length scales which can be revealed by using e.g. nuclear magnetic resonance (NMR) spectroscopy or neutron and synchrotron x-ray diffraction methods. $^{27}$Al NMR and diffraction measurements show that glasses in the (CaO)$_{x}$(Al$_{2}$O$_{3}$)$_{1-x}$ system with $x>0.5$ are composed of Al coordinated by four oxygen atoms, while more Al$_{2}$O$_{3}$-rich glasses contain five- and six-fold coordinated units \cite{McMillan96,Hannon00,Benmore03,Drewitt12b}. {\it In situ} $^{27}$Al NMR experiments of liquid aluminates similarly reveal larger populations of highly coordinated Al-units with increasing Al$_{2}$O$_{3}$ fractions, however motional averaging prevents the identification of individual coordination environments \cite{Poe93,Poe94,Massiot95}. Neutron and synchrotron x-ray diffraction measurements of these liquids reveal up to 20\,\% five-fold Al coordinated units at $x=0.33$, with aluminium tetrahedra becoming increasingly dominant towards higher CaO fractions \cite{Weber03,Hennet07a,Mei08b,Cristiglio10,Drewitt11,Drewitt12a,Drewitt12b,Drewitt17}. Diffraction experiments provide information on the atomic-scale structure of liquids in the form of the pair-distribution function $G(r)$, which provides a measure of the probability of finding two atoms a distance $r$ apart. For a system comprising $n$ different chemical species, $G(r)$ comprises a weighted sum of $n(n+1)/2$ overlapping partial pair-distribution functions $g_{\alpha\beta}(r)$. This complexity can make it difficult to unambiguously interpret the experimental data. However, measurements made using both neutron and synchrotron x-ray diffraction are highly complementary: while x-rays are sensitive to elements with high atomic numbers, neutrons are sensitive to lighter elements such as oxygen.  Furthermore, since the scattering power of neutrons varies between isotopes of the same element, the method of neutron diffraction with isotope substitution (NDIS) can be used to provide site-specific information. In NDIS, diffraction measurements are made for two or more samples that are identical in every respect, except for the isotopic enrichment of one or more of the chemical species. A subtraction of the measured diffraction patterns allows the complex overlapping $g_{\alpha\beta}(r)$ functions involving the substituted elements to be unraveled, providing highly-detailed information on liquid structure \cite{Drewitt12b,Drewitt17}. The results of these scattering techniques in turn provide a rigorous test of the efficacy of the interaction models used in molecular dynamics (MD) simulations, enabling a full picture of the liquid structural correlations to be obtained \cite{Drewitt11,Drewitt12b,Drewitt17}. 

In this paper we present new time-resolved synchrotron x-ray diffraction measurements, and review recent progress in neutron diffraction with isotope substitution experiments combined with molecular dynamics simulations using an aspherical ion model, to provide detailed insight into the structural transformations which take place during vitrification of calcium aluminate liquids and the atomic-scale configurational constraints on their glass-forming ability. 

\section{Methods and Results}

\subsection{Aerodynamic levitation for liquid diffraction}
\begin{figure}
\centering
\includegraphics[width=0.48\textwidth]{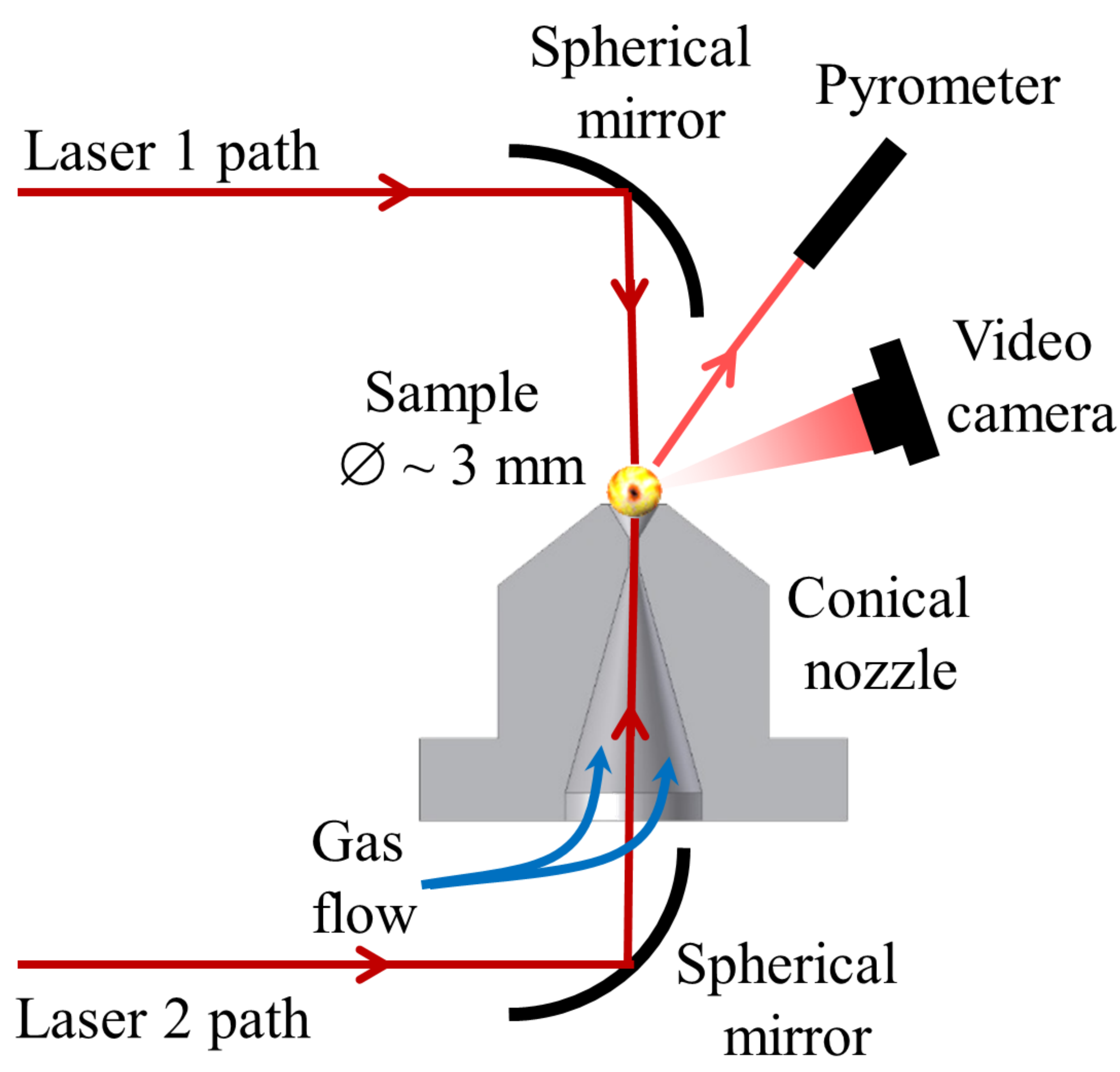}
\caption{A schematic of the aerodynamic levitation method showing a convergent-divergent levitation nozzle in cross-section with infra-red lasers focused onto the sample from above and below. A regulated gas-flow counteracts gravity to levitate high-temperature liquids in the upper cone of the nozzle. The sample is monitored and temperature measured using a video camera and pyrometer, respectively.\label{Fig3-Lev}}
\end{figure}
{\it In situ} liquid diffraction measurements of aluminates are challenging due to their high-melting temperatures ($>1500$\,$^{\circ}$C) at which conventional furnaces present a high risk of chemical reaction with a sample. This problem can be overcome by containerless processing, where common methods include levitation using an acoustic \cite{Brandt01,Marzo17}, aerodynamic \cite{Krishnan97}, electromagnetic \cite{Jacobs96}, electrostatic \cite{Paradis08}, or optical field \cite{Neuman04}. Of these techniques, aerodynamic levitation has been most widely used for the synthesis and characterisation of oxide liquids and glasses \cite{Price10,Hennet11a,Benmore17}. In this method, a sample is placed in the upper cone of a convergent-divergent conical nozzle, which channels a regulated gas (e.g. Ar, N$_{2}$) flow upwards to counteract gravity (figure \ref{Fig3-Lev}). The sample is thus levitated without contact just above the nozzle, and heating can be achieved by using lasers focused on to the sample from above and below. In this way, high-temperature oxide liquids and metastable supercooled states can be stably levitated indefinitely \cite{Price10}. Spherical glasses of diameter 1-4\,mm can also be synthesised by rapidly quenching a levitated liquid by abruptly cutting the laser power. The  method is ideally suited for {\it in situ} x-ray and neutron scattering measurements of oxide liquids as aerodynamic levitation devices can be readily transported and integrated into central facility beamlines and, with no container, clean data sets are obtained, thereby reducing the complexity of the data analysis. This enables the application of advanced techniques, such as time-resolved synchrotron x-ray diffraction or neutron diffraction with isotope substitution, to reveal detailed insight into the liquid and glass structures.

\subsection{Molecular dynamics simulations}
MD simulations can provide a full-scale model of the structural correlations of liquids. First-principles MD, based on a quantum mechanical method such as density functional theory to derive interatomic forces, can calculate properties of liquids to a high accuracy. However, the computational cost is very high and involves small volumes (a few 100 atoms) and timescales (a few tens of ps), such that the dynamical processes of glass formation are not well represented. Classical MD, on-the-other-hand, uses parameterised force fields and has the capacity to study larger systems with longer run times. However, classical potentials can be of limited accuracy.

The simulation results presented in this paper were generated using a set of advanced ionic interaction (AIM) potentials derived for the Ca-Mg-Al-Si-O (CMAS) system and parameterised by fitting the predicted forces and multipoles to first-principles calculations to account for dipole polarisation effects and ion shape deformations \cite{Jahn07}. The interaction potential $V$ is constructed from four components \cite{Drewitt11}
\begin{equation}
V=V^{qq}+V^{\rm{disp}}+V^{\rm{rep}}+V^{\rm{pol}},
\end{equation}
The charge-charge ($V^{qq}$) and dispersion ($V^{\rm{disp}}$) interactions are purely pairwise additive:
\begin{equation}
V^{qq}=\sum_{i\le j}\frac{q^iq^j}{r_{ij}},
\end{equation}
\begin{equation}
V^{\rm{disp}} = -\sum_{i\le j} [f_6^{ij}(r_{ij})C_6^{ij}/r_{ij}^6
                       +  f_8^{ij}(r_{ij})C_8^{ij}/r_{ij}^8],
\end{equation}
$q^i$ are the formal charges of ions $i$ (+3 for Al, +2 for Ca and -2 for O).
$C_6^{ij}$ and $C_8^{ij}$ are the dipole-dipole and dipole-quadrupole dispersion coefficients, and $f_n^{ij}$ are Tang-Toennies dispersion damping functions ($X=6,8$; $N_6=6$; $N_8=8$; $c_X=1.0$) \cite{Tang84}
\begin{equation}
f_X(r^{ij}) = 1-c_Xe^{-b_Xr^{ij}}\sum_{k=0}^{N_X}\frac{(b_Xr^{ij})^k}{k!},
\end{equation}
which describe short-range corrections to the asymptotic dispersion terms.

$V^{\rm{rep}}$ describes the overlap repulsion interaction
\begin{eqnarray}
V^{\rm{rep}} &=& \sum_{i\in O,j\in Ca,Al} [A^{-+}e^{-a^{-+}\rho^{ij}} +
          B^{-+}e^{-b^{-+}\rho^{ij}} + \nonumber \\
         &&  C^{-+}e^{-c^{-+} r_{ij}}] 
          + \sum_{i,j\in O} A^{--}e^{-a^{--}r_{ij}} + \nonumber \\
         && \sum_{i\in O} [D(e^{\beta\delta\sigma^i}
          +e^{-\beta\delta\sigma^i}) +
          (e^{\zeta^2\mid{\bf \nu}^i\mid^2} - 1) \nonumber \\
         && + (e^{\eta^2\mid{\bf \kappa}^i\mid^2} - 1)]
\label{eqn:ov}
\end{eqnarray}
with
\begin{equation}
\rho^{ij} = r_{ij} - \delta\sigma^i -
S_{\alpha}^{(1)}\nu_{\alpha}^i -
S_{\alpha\beta}^{(2)}\kappa_{\alpha\beta}^i,
\end{equation}
and summation of repeated indexes is implied.
The variable $\delta\sigma^i$ characterizes the deviation of the radius of
oxide anion $i$ from its default value, \{$\nu_{\alpha}^i$\} are a set
of three variables describing the Cartesian components of
a dipolar distortion of the ion, and \{$\kappa_{\alpha\beta}^i$\} are a set
of five independent variables describing the corresponding quadrupolar shape
distortions ($\mid\kappa\mid^2 =
\kappa_{xx}^2+\kappa_{yy}^2+\kappa_{zz}^2+2(\kappa_{xy}^2+\kappa_{xz}^2
+\kappa_{yz}^2)$ with a traceless matrix $\kappa$).
$S_{\alpha}^{(1)}=r_{ij,\alpha}/r_{ij}$ and $S_{\alpha\beta}^{(2)}=
3r_{ij,\alpha}r_{ij,\beta}/r_{ij}^2 - \delta_{\alpha\beta}$ are interaction
tensors.
The last set of sums are self-energy terms representing the energy
required to deform the anion charge density,
with $\beta$, $\zeta$ and $\eta$ as effective force constants.
The extent of each ion's distortion is determined at each molecular
dynamics time-step by energy minimization.

The polarization part of the potential incorporates dipolar
and quadrupolar contributions \cite{Wilson96},
\begin{eqnarray}
V^{\rm{pol}}&=& \sum_{i\le j}
        \left( (f_{D}^{ij}(r_{ij})q^i\mu^j_{\alpha}-f_{D}^{ji}(r_{ij})q^j\mu^i_{\alpha})T_{\alpha}^{(1)} 
 \right. \nonumber \\ &&
     +(f_{Q}^{ij}(r_{ij})\frac{q^i\theta^j_{\alpha\beta}}{3}
     + f_{Q}^{ji}(r_{ij})\frac{\theta^i_{\alpha\beta}q^j}{3}
       -\mu^i_{\alpha}\mu^j_{\beta})T_{\alpha\beta}^{(2)}
\nonumber \\ && \left.
     +(\frac{\mu^i_{\alpha}\theta^j_{\beta\gamma}}{3}
     +\frac{\theta^i_{\alpha\beta}\mu^j_{\gamma}}{3})T_{\alpha\beta\gamma}^{(3)}
     + \frac{\theta^i_{\alpha\beta}\theta^j_{\gamma\delta}}{9}
       T_{\alpha\beta\gamma\delta}^{(4)})\right) \nonumber \\ &&
     + \sum_{i \in O,Ca} \left(\frac{1}{2\alpha^i}\mid\vec\mu^{i}\mid^2 +
                  \frac{1}{6C^i}\theta^i_{\alpha\beta}\theta^i_{\alpha\beta} \right)
\label{vpol}
\end{eqnarray}
where
$\alpha^i$ and $C^i$ are the dipole and quadrupole polarizabilities. Only the oxygen and calcium ions are considered polarizable.
$T_{\alpha\beta\gamma\delta} =
\nabla_{\alpha}\nabla_{\beta}\nabla_{\gamma}\nabla_{\delta}...\frac{1}{r_{ij}}$
are the multipole interaction tensors \cite{Stone96}.
The instantaneous values of these moments are obtained by minimization of this expression. The charge-dipole and charge-quadrupole asymptotic functions include terms are damped at short distances by Tang-Toennies functions \cite{Tang84} with $X=(D,Q)$, $N_D=4$ and $N_Q=6$. Short-range damping of the anion-anion functions is neglected. The parameters $b_D$ and $b_Q$ determine the range at which the overlap of the charge densities affects the induced multipoles, the parameters $c_D$ and $c_Q$ determine the strength of the ion response to this effect.
The potential parameters are listed in table \ref{Table1}.

\begin{table}
  \caption{AIM potential parameters (in atomic units, from \cite{Drewitt11}).}\label{Table1}
  \begin{indented}
  \item[]\begin{tabular}{llllll}
  \br
               & O-O   & Al-O  & Ca-O  & (O,Al)-Ca  \\
\mr
$A^{ij}$       & 1068.0& 18.149& 40.058\\
$a^{ij}$       & 2.6658& 1.4101& 1.5035\\
$B^{ij}$       &       & 51319.& 50626.\\
$b^{ij}$       &       & 3.8406& 3.5024\\
$C^{ij}$       &       & 6283.5& 6283.5\\
$c^{ij}$       &       & 4.2435& 4.2435\\
\\
$b_{D}^{ij}$   &       & 2.2886& 1.8297& 3.50\\
$c_{D}^{ij}$   &       & 2.3836& 2.3592& 1.00\\
$b_{Q}^{ij}$   &       & 2.1318& 1.0711& 1.09\\
$c_{Q}^{ij}$   &       & 1.2508& 1.0000& 1.00\\
\\
$C_6^{ij}$     & 44.372& 2.1793& 2.1793\\
$C_8^{ij}$     & 853.29& 25.305& 25.305\\
$b_{\rm{disp}}^{ij}$& 1.4385& 2.2057& 2.2057\\
\mr
$D$            &0.49566\\
$\beta$        & 1.2325\\
$\zeta$        &0.89219\\
$\eta$         & 4.3646&       &       & Ca\\
$\alpha^i$     & 8.7671&       &       & 3.50\\
$C^i$          &11.5124&       &       & 4.98\\

  \br
  \end{tabular}
  \end{indented}
\end{table}
By comparison to purely pairwise classical interaction potentials, the CMAS AIM potentials have a higher degree of transferability between different systems.
Despite the fact that there are no explicit three-body or angular terms in the AIM potential, the self-consistent evaluation of the ion deformability ($V^{\rm{rep}}$) and polarizability ($V^{\rm{pol}}$) parts of the potential energy in each step of the MD simulation leads to an implicit account of many-body interactions.

The AIM potentials have been applied successfully to study the properties of MgO-Al$_{2}$O$_{3}$ \cite{Jahn08} and MgO-SiO$_{2}$ \cite{Adjaoud08} liquids, as well as the CaO-Al$_{2}$O$_{3}$ system studied and reviewed here \cite{Drewitt11,Drewitt12b,Drewitt15,Drewitt17}. The simulation cells contained 1512 ions for the CaAl$_2$O$_4$ (CA) and 1892 ions for the Ca$_3$Al$_2$O$_6$ (C3A) system.The liquid-state AIM-MD simulations were performed with a time step of 1\,fs at constant temperature 2230\,$^{\circ}$C. The first 50 ps of the simulations were used to equilibrate the melts at constant pressure (1 bar). This was followed by 100 ps production runs at constant volume \cite{Drewitt11}. The structure of CA glass at 30\,$^{\circ}$C was obtained from the melt employing a constant pressure simulation with a quench rate of 10$^{12}$\,Ks$^{-1}$ \cite{Drewitt12b}. In this work, we also present the results of new AIM-MD simulations of C3A glass using the same quench rate.

\subsection{Time-resolved synchrotron x-ray diffraction}
\begin{figure*}
\centering
\includegraphics[width=1\textwidth]{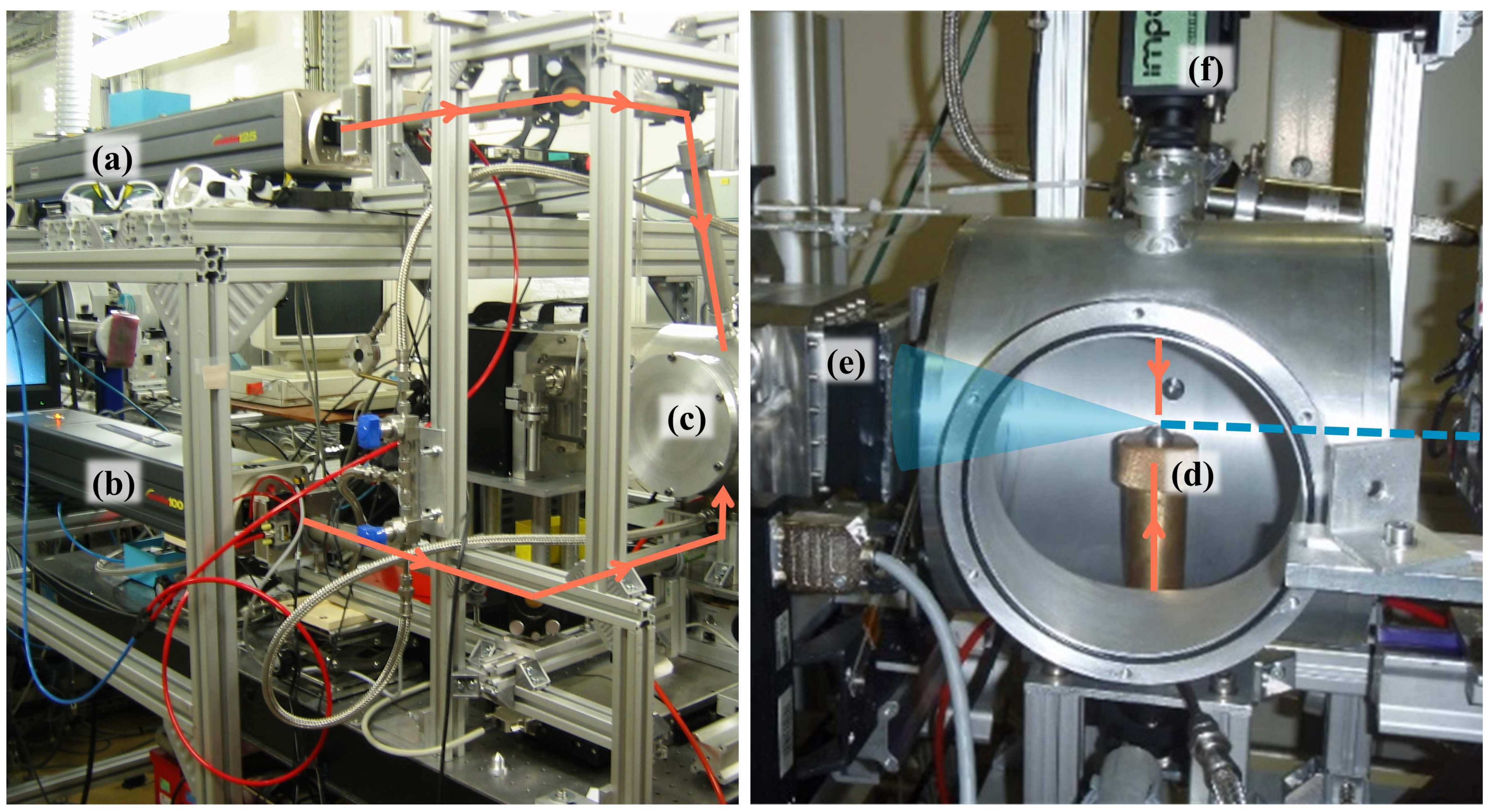}
\caption{Experimental setup at beamline ID11 at the European Synchrotron Radiation Facility (ESRF), France.  The photograph on the left shows the laser path from two CO$_{2}$ lasers (a) and (b), indicated by the light red arrows and enclosed by metallic tubing for safety, reflected by flat Cu mirrors and focused onto the sample position within the levitation chamber (c) from above and below using spherical Cu mirrors. The right hand photograph shows the x-rays indicated by the blue dashed line incident on the laser heated sample on the levitation nozzle (d) with the diffraction recorded on the Frelon CCD detector (e). Temperature is recorded using a pyrometer (f) and the sample is monitored remotely during a diffraction experiment using a video camera (not shown).
\label{Fig4-ID11}}
\end{figure*}
\begin{figure}
\centering
\includegraphics[width=0.48\textwidth]{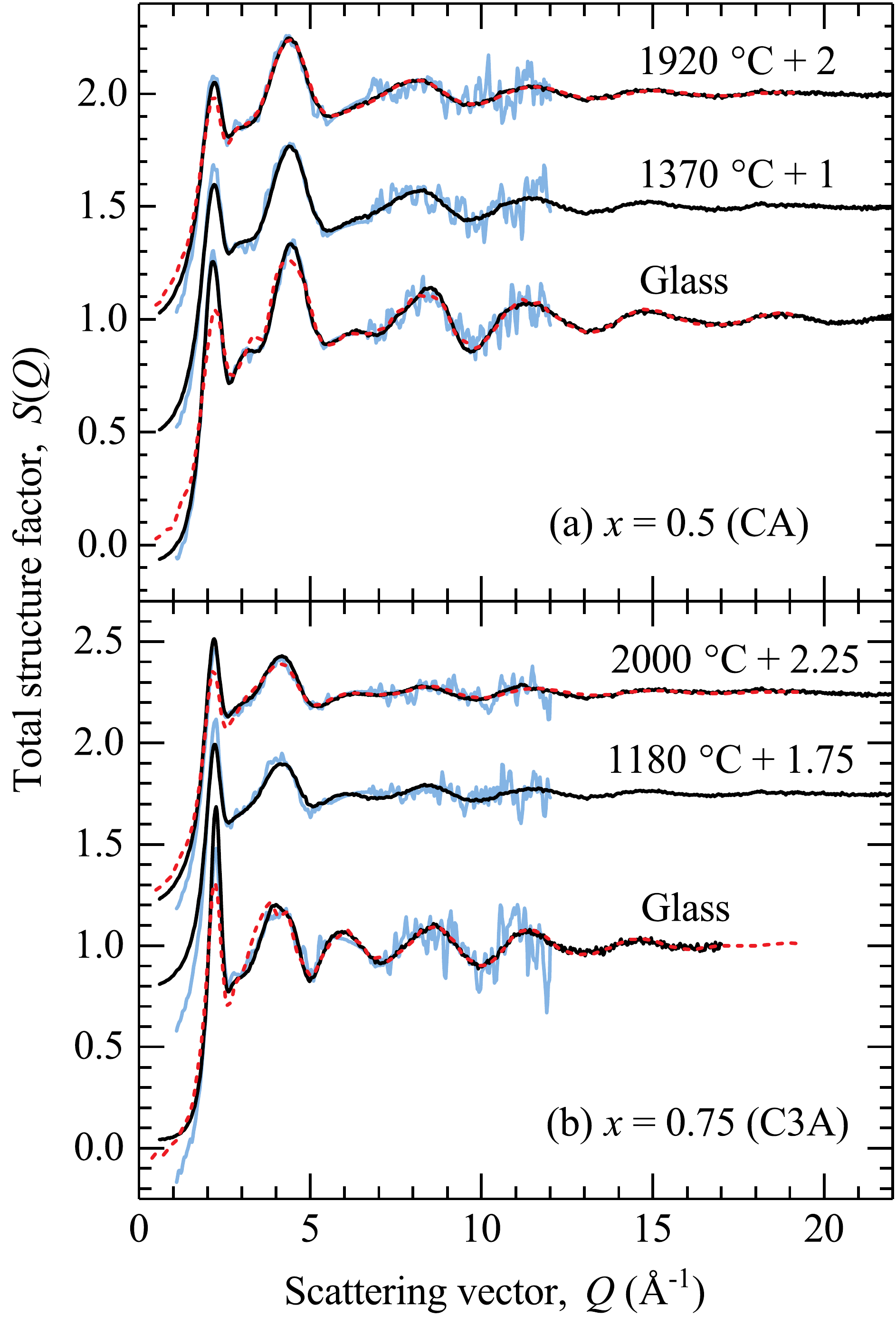}
\caption{Total structure factors $S(Q)$ for the (CaO)$_{x}$(Al$_{2}$O$_{3}$)$_{1-x}$ liquids (a) $x=0.5$ (CA) and (b) $x=0.75$ (C3A), their supercooled liquids, and glasses as obtained from 60\,s SXRD measurements (dark black curves) and time-resolved 30\,ms acquisitions (light blue curves). The red dashed curves are the $S(Q)$ functions generated from the AIM-MD simulations obtained for CA and C3A liquids at $2230^{\circ}$C \cite{Drewitt11}, and for the CA \cite{Drewitt12b} and C3A glasses at $30^{\circ}$C. \label{Fig5-SXRD-SQ}}
\end{figure}
\begin{figure}
\centering
\includegraphics[width=0.48\textwidth]{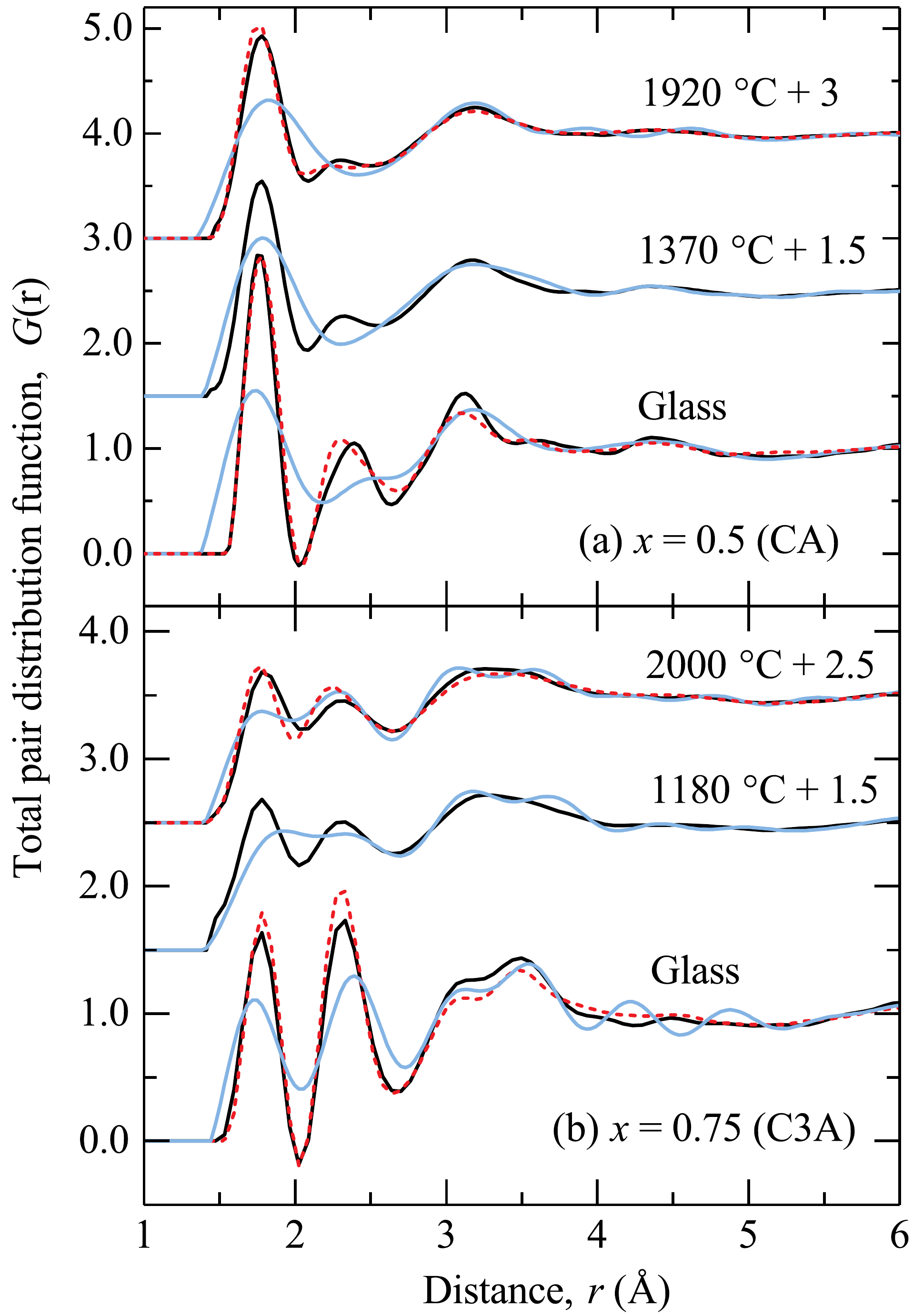}
\caption{Total pair distribution functions $G(r)$ for the (CaO)$_{x}$(Al$_{2}$O$_{3}$)$_{1-x}$ liquids (a) $x=0.5$ (CA) and (b) $x=0.75$ (C3A), as obtained by Fourier transforming the corresponding $S(Q)$ functions shown in figure \ref{Fig5-SXRD-SQ} from the static (dark black curves) and time-resolved (light blue curves) SXRD measurements and AIM-MD simulations \cite{Drewitt11,Drewitt12b} (red dashed curves). \label{Fig6-SXRD-Gr}}
\end{figure}
\begin{table*}
\caption{Real-space peak positions $r_{\rm{AlO}}$, $r_{\rm{CaO}}$, $r_{\rm{CaCa}}$, and coordination numbers $\bar{n}_{\rm{Al}}^{\rm{O}}$, $\bar{n}_{\rm{Ca}}^{\rm{O}}$, $\bar{n}_{\rm{Ca}}^{\rm{Ca}}$ for the (CaO)$_{x}$(Al$_{2}$O$_{3}$)$_{1-x}$ liquids with $x=0.5$ (CA) and $x=0.75$ (C3A), from synchrotron x-ray diffraction (SXRD), neutron diffraction (ND), and neutron diffraction with isotope substitution (NDIS) measurements, or aspherical ion model molecular dynamics (AIM-MD) simulations.}\label{Table2}
\begin{tabular}{llllllllll}
\toprule
$x$ & $T$ ($^{\circ}$C) & Technique &
$r_{\rm{AlO}}\,(\rm{\AA})$ & $r_{\rm{CaO}}\,(\rm{\AA})$ & $r_{\rm{CaCa}}\,(\rm{\AA})$ & $\bar{n}_{\rm{Al}}^{\rm{O}}$ & $\bar{n}_{\rm{Ca}}^{\rm{O}}$ & $\bar{n}_{\rm{Ca}}^{\rm{Ca}}$ & Reference
\\\midrule
0.5 & 2230 & AIM-MD & 1.75(1) & 2.29(1) & 3.86(2) & 4.13(5) &  6.2(1) & 5.1(1) & \cite{Drewitt11}\\ 
0.5 & 1920 & SXRD & 1.78(1) & 2.30(2) & -- & 4.24(5) & 5.5(1) & -- & This work\\
0.5 & 1700 & NDIS & 1.77(1) & 2.30(1) & -- & 4.20(4) & 6.0(2) & -- & \cite{Drewitt12b}\\
0.5 & 1370 & SXRD & 1.78(1) & 2.31(2) & -- & 4.14(5) & 6.0(2) & -- & This work \\
0.5 & 30 & AIM-MD & 1.76(1) & 2.34(1) & 3.81(5) & 4.11(5) & 6.4(1) & 4.9(1) & \cite{Drewitt12b}\\
0.5 & 20 & SXRD & 1.76(1) & 2.39(1) & -- & 4.30(5) & 6.0(1) & --& This work \\
0.5 & 20 & NDIS & 1.75(1) & 2.35(1) & 3.59(2) \& 4.41(5) & 4.04(3) & 6.4(2) & 5.4(1) &  \cite{Drewitt12b} \\
0.75 & 2230 & AIM-MD & 1.75(1) & 2.26(1) & 3.65(3) & 4.04(5) & 5.6(1) & 8.5(1) & \cite{Drewitt11}\\ 
0.75 & 2000 & SXRD & 1.79(1) & 2.31(2) & -- & 4.28(5) & 4.4(1) & -- & This work\\
0.75 & 1800 & NDIS & 1.75(1) & 2.26(2) & 3.62(4) & 4.1(1) & 5.7(2) & 7.9(5) & \cite{Drewitt17}\\
0.75 & 1180 & SXRD & 1.78(1) & 2.31(2) & -- & 4.26(5) & 4.6(1) & --& This work\\
0.75 & 30 & AIM-MD & 1.76(1) & 2.30(1) & 3.48(2) \& 3.80(5) & 4.01(1) & 5.5(1) & 8.6(1) & This work\\
0.75 & 20 & SXRD & 1.75(1) & 2.35(2) & -- & 4.3(1) & 4.8(1) & -- & This work\\
0.75 & 20 & ND & 1.74(1) &  2.32(3) & -- & 4.2(1) & 4.3(1) & -- & This work\\
\bottomrule
\end{tabular}
\end{table*}
\begin{figure*}
\centering
\includegraphics[width=0.98\textwidth]{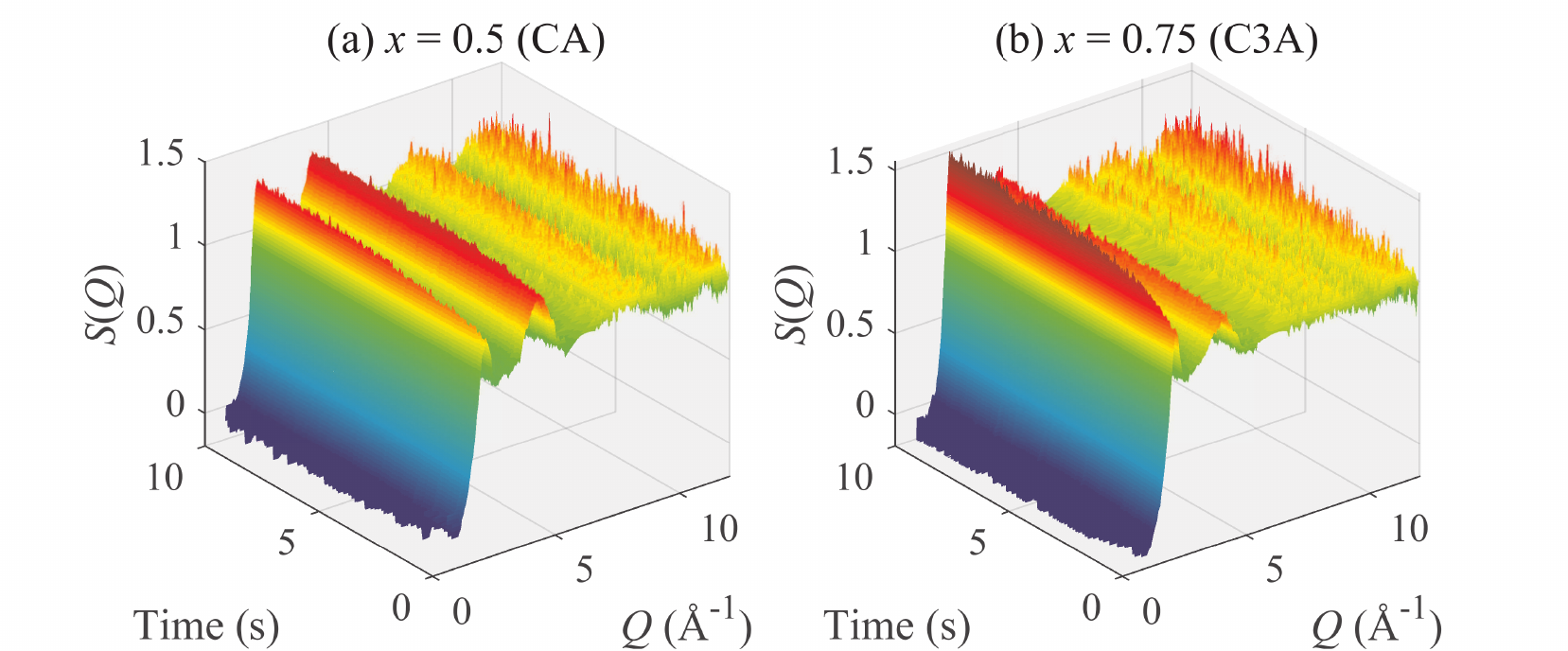}
\caption{Time-resolved total structure factors $S(Q)$ for the (CaO)$_{x}$(Al$_{2}$O$_{3}$)$_{1-x}$ liquids (a) $x=0.5$ (CA) and (b) $x=0.75$ (C3A) during vitrification from the high-temperature liquid to the ambient-temperature glass.\label{Fig7-TRSXRD}}
\end{figure*}
\begin{figure}
\centering
\includegraphics[width=0.48\textwidth]{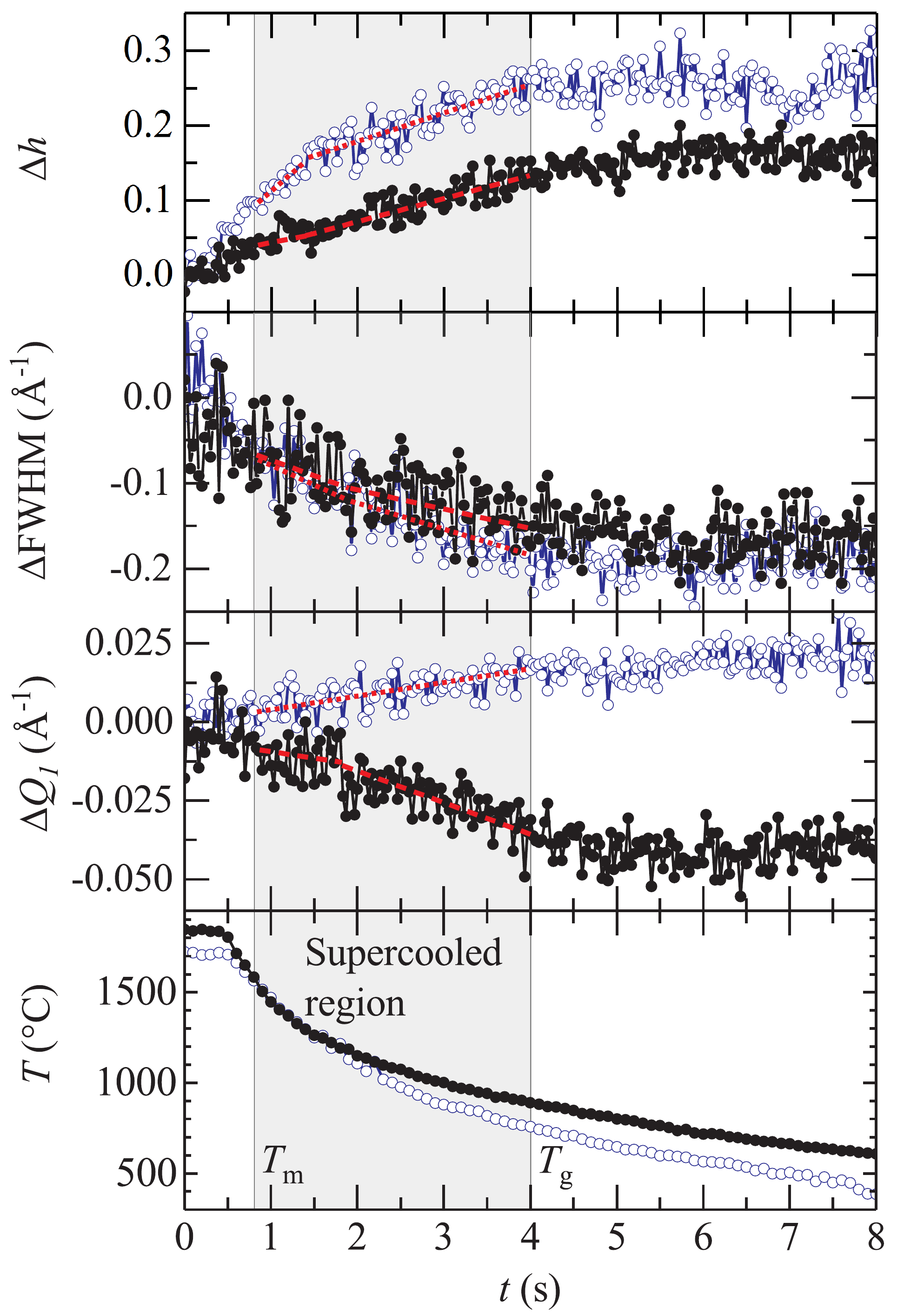}
\caption{Cooling curves for liquid CA (black closed circles) and C3A (blue open circles) showing temperature $T$ as a function of time $t$, together with the relative changes in position $\Delta Q_{1}$, full width at half maximum $\Delta$FWHM, and height $\Delta h$ of the first peak in $S(Q)$. Red dashed lines are guides for the eye. The supercooled region between the melting $T_{\rm{m}}$ and glass transition $T_{\rm{g}}$ temperatures of the liquids is approximately indicated by the shaded region.}\label{Fig8-Peakfit}
\end{figure}

High-flux synchrotron sources enable x-ray scattering measurements of liquids and glasses to be collected with millisecond time-resolution using fast detectors. Combined with aerodynamic levitation with laser heating, time-resolved diffraction measurements can be made for liquid oxides to follow the development of structural changes as these liquids are supercooled through the glass transition into the solid state \cite{Hennet05,Hennet07b,Hennet08,Bytchkov10,Hennet11a,Hennet11b,Skinner13b}. Here we report new time-resolved synchrotron x-ray diffraction (SXRD) measurements to track glass formation in (CaO)$_{x}$(Al$_{2}$O$_{3}$)$_{1-x}$ liquids with $x=0.5$ (CaAl$_{2}$O$_{4}$ denoted CA) and $x=0.75$ (Ca$_{3}$Al$_{2}$O$_{6}$ denoted C3A) using the aerodynamic levitation with laser heating setup  \cite{Hennet08} at beamline ID11 at the European Radiation Facility (ESRF), France (figure \ref{Fig4-ID11}). The sample was levitated by an Ar\,$+3$\,\%\,O$_{2}$ gas flow and melted using two 125\,W CO$_{2}$ lasers incident from above and below the sample. Temperatures in the range 350-3000\,$^{\circ}$C were measured using an optical pyrometer.  Diffraction by high-energy x-rays (100.456\,keV) was recorded using the Fast Readout Low Noise (FReLoN) 2-dimensional charge-coupled device (CCD) detector\cite{Labiche07}. The camera uses an ATMEL chip 7899M, which operating in full-frame-transfer mode, has an active image zone of $2048\times2048$ pixels of size 14\,$\mu\rm{m}^{2}$.  In frame-transfer mode the active image zone is reduced to $2048\times1024$ pixels, where the remainder of the chip is used as a temporary memory buffer to store the previous image for simultaneous readout and data collection to enable consecutive fast data acquisitions. 

Static diffraction measurements were made at ID11 with the camera in full-frame mode with an acquisition time of 60\,s for the CA and C3A liquids several hundreds of degrees above their melting points $T_{\rm{m}}$ ($1605\,^{\circ}$C for CA, $1541\,^{\circ}$C for C3A \cite{Nurse65}), their supercooled liquids several hundreds of degrees below their melting points but above $T_{g}$  ($905\,^{\circ}$C for CA \cite{Poe94,Massiot95}, $771\,^{\circ}$C for C3A \cite{Neuville19}) and the glasses at ambient temperature. The static diffraction measurement for C3A glass was made at the DIFFABS beamline at the Soleil Synchrotron, France, using an 18\,keV incident x-ray beam with the scattering signal scanned using a scintillation detector with 3\,s counting time per 0.2$^{\circ}$ step over an angular range of 2-140$^{\circ}$. Time-resolved diffraction measurements were made at ID11 with the camera in frame-transfer mode with successive 30\,ms acquisitions recorded as the high-temperature liquids were quenched, by abruptly switching off the laser-power, and supercooled through the glass transition T$_{\rm{g}}$ into the solid state. The data were analysed using the procedure detailed in reference \citenum{Drewitt11} to obtain the total structure factors, defined at high-energies far from an absorption edge by 
\begin{equation}\label{SQ-eq}
S(Q)=\sum_{\alpha=1}^{n}\sum_{\beta=1}^{n}\frac{c_{\alpha}c_{\beta}f_{\alpha}(Q)f_{\beta}(Q)}{\left[\sum_{\alpha}c_{\alpha}f_{\alpha}(Q)\right]^{2}}[S_{\alpha\beta}(Q)-1],
\end{equation}
where $n=3$ is the number of chemical species $\alpha$ or $\beta$ (Al, Ca, O),  $Q$ is the magnitude of the scattering vector, $c_{\alpha}$ and $c_{\beta}$ are the atomic concentrations, $f_{\alpha}(Q)$ and $f_{\beta}(Q)$ are the atomic form factors, and $S_{\alpha\beta}(Q)$ is a Faber-Ziman \cite{Faber65} partial structure factor. The $S(Q)$ functions measured by static diffraction are shown in figure \ref{Fig5-SXRD-SQ} together with selected time-resolved measurements at comparable temperatures. The time-resolved $S(Q)$ functions are in good overall agreement with the static diffraction measurements and exhibit improved counting statistics and larger maximum scattering vector $Q_{\rm{max}}$ compared to previous work, despite a factor of 3 shorter acquisition times \cite{Hennet07b}. The corresponding total pair distribution functions shown in figure \ref{Fig6-SXRD-Gr} were calculated by the Fourier transformation 
\begin{equation}\label{Gr-FT}
G(r)-1=\frac{1}{2\pi^{2}rn_{0}}\int_{0}^{Q_{\rm{max}}}Q[S(Q)-1]\frac{\sin Qr}{r}M(Q)\rm{d}Q,
\end{equation}
where $n_{0}$ is the atomic number density and $M(Q)$ is a cosine modification function \cite{Drewitt13} used to reduce the termination ripples generated as a result of the finite accessible $Q_{\rm{max}}$. The time-resolved $G(r)$ functions are considerably broadened as a result of the smaller $Q_{\rm{max}}=12\,\rm{\AA}^{-1}$ compared to $22\,\rm{\AA}^{-1}$ for the static diffraction experiments, which leads to a loss in information on the local structural ordering that is encoded in the high-$Q$ oscillations. The $S(Q)$ and $G(r)$ functions generated from the AIM-MD simulations are in very good agreement with the experimental measurements, although the height of the first peak in $S(Q)$ for the simulated glasses is smaller than in the experimental measurements. The full range of $S(Q)$ functions obtained from the time-resolved measurements of liquid CA and C3A are shown in figure \ref{Fig7-TRSXRD}.  The main peaks and diffuse high-$Q$ oscillations are enhanced in the glass $S(Q)$ functions, compared to the liquids, due to the higher degree of thermal motion and overall structural disorder in the liquid-state. In both CA and C3A liquids, the first peak in $S(Q)$ at $Q_{1}=2.19(2)\,\rm{\AA}^{-1}$ experiences a substantial development in height on vitrification. The position of the first peak in $S(Q)$ can be used to classify ordering in liquids and glasses \cite{Price89}. The values $Q_{1}$ and $Q_{1}r_{1}=3.90(1)$ (where $r_{1}=1.78(1)\,\rm{\AA}$ is the position of the first peak in $G(r)$) are at the upper limit for attribution as a so-called first sharp diffraction peak (FSDP), indicative of ordering of cation-centred polyhedra on intermediate length scales \cite{Price89,Elliot91}. However, the position is consistent with the linear increase in FSDP position observed in calcium aluminosilicate glasses with reducing SiO$_{2}$ concentration \cite{Petkov98,Hennet07a}. The AIM-MD simulations show that this peak arises predominantly from $S_{\rm{AlCa}}(Q)$ and $S_{\rm{CaCa}}(Q)$, and is sharper in C3A due to the higher concentration of Ca and hence greater influence of these cation-cation correlations \cite{Drewitt11}. In $G(r)$, the first peak at $\sim1.78$\,$\rm{\AA}$ is attributed to the nearest neighbour Al-O bond, the second peak at $\sim2.3$\,$\rm{\AA}$ arises from Ca-O correlations, and the third peak at $\sim3.2\,\rm{\AA}$ results from a combination of Al-Al, Al-Ca, and O-O correlations \cite{Drewitt11}. For C3A, the Al-O peak has reduced height, and the Ca-O peak is greatly enhanced, compared to the CA measurement, consistent with the increase in CaO fraction. On vitrification, both the Al-O and Ca-O peak experience a significant increase in height and change in position to $r_{\rm{AlO}}=1.75\,\rm{\AA}$ and $r_{\rm{CaO}}=2.35\,\rm{\AA}$, consistent with a reduction and increase in the Al-O and Ca-O coordination numbers, respectively. Real-space peak positions and average coordination numbers are listed in table \ref{Table2}. Coordination numbers were obtained by integrating over a relevant peak in the real-space Fourier transform of a modified $S(Q)$ function in which the $Q$-dependent weighting factors applied to the partial structure factors of interest have been eliminated \cite{Drewitt11,Zeidler09}.

The cooling curve for the time-resolved experiments is shown in figure \ref{Fig8-Peakfit}, together with the relative changes in position $\Delta Q_{1}$, full width at half maximum $\Delta \rm{FWHM}$, and height $\Delta h$ of a Lorentzian function fitted to the first peak in $S(Q)$ in each 30\,ms acquisition. Cooling from the respective $T_{\rm{m}}$ to $T_{\rm{g}}$ took approximately 3.2\,s for each liquid, such that $\sim96\times30$\,ms SXRD measurements were recorded in the supercooled regions during glass formation. The cooling rate d$T/$d$t$ decreases as a function of time, with two distinct cooling regimes in the supercooled region, marked by an inflection in the vicinity of the dynamical cross-over temperature at $\simeq1.25T_{\rm{g}}$, with d$T/$d$t=325$\,Ks$^{-1}$ (CA) and 340\,Ks$^{-1}$ (C3A) at $T>1.25$\,$T_{\rm{g}}$, or d$T/$d$t=125$\,Ks$^{-1}$ (CA) and 142\,Ks$^{-1}$ (C3A) at $T_{\rm{g}}<T<1.25$\,$T_{\rm{g}}$. Considering liquid CA, the first peak in $S(Q)$ undergoes a progressive increase in height and reduction in width consistent with previous work \cite{Hennet07b}.  These changes are accompanied by a continuous shift in peak position to lower $Q$-values. As for the cooling curves, an inflection is also observed in these peak parameters at $\sim1.25\,T_{\rm{g}}$, indicative of configurational modifications taking place close to the dynamical crossover temperature. Considering the preceding discussion on the FSDP, these changes are indicative of a progressive ordering of cation-centred polyhedra on an intermediate-range length scale as the temperature decreases. At $T_{\rm{g}}$, no further structural evolution is observed, consistent with the freezing of the supercooled liquid into the solid glassy state. For liquid C3A, the first peak in $S(Q)$ experiences a similar height and width evolution to liquid CA with a slightly more pronounced $\Delta h$. However, in contrast to liquid CA, the position $Q_{1}$ experiences a small but significant shift to higher $Q$-values on vitrification. Examination of the AIM-MD simulation results reveal the peak shift in CA to lower-$Q$ values arises from a complex superposition of the reciprocal-space partial correlations in this region, whereas the shift to larger $Q$ observed on vitrification of liquid C3A is attributed $S_{\rm{CaCa}}(Q)$.
\subsection{Neutron diffraction with isotope substitution}
\begin{figure*}[htpb]
\centering
\includegraphics[width=1\textwidth]{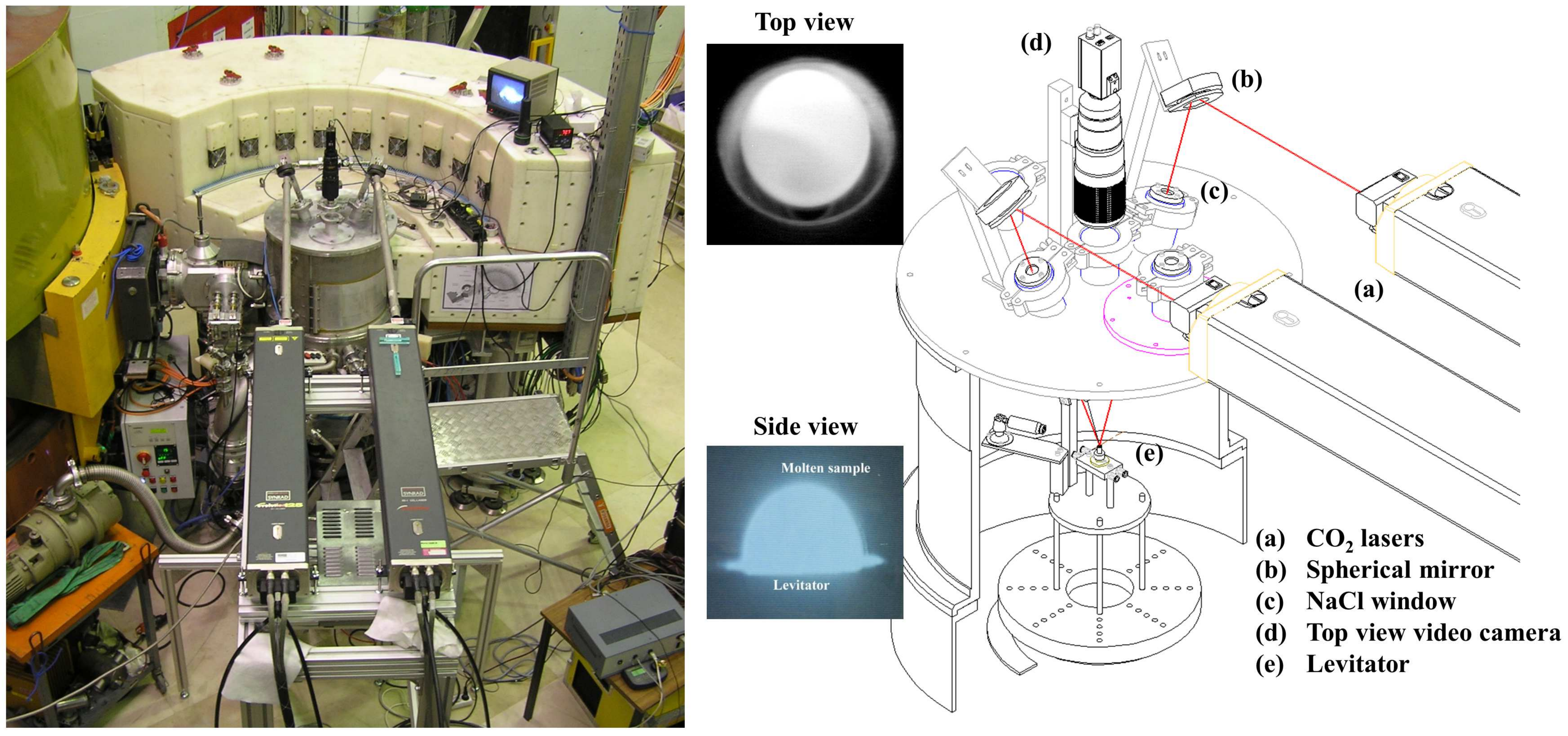}
\caption{Photograph and schematic of the aerodynamic levitation with laser-heating setup \cite{Hennet06} installed at the D4c neutron diffractometer \cite{Fischer02} at the Institut Laue-Langevin, France. Spherical samples of diameter $\sim\,3$\,mm are levitated on a flow of gas using a conical vanadium nozzle, which scatters neutrons almost entirely incoherently. Monochromatic neutrons of wavelength $0.4979(1)\,\rm{\AA}$ (incident from the left in the photograph) are collimated to around 5\,mm vertically by 10\,mm horizontally by neutron-absorbing $^{10}$B$_{4}$C flags positioned within the sample belljar close to the levitation nozzle to mask it from the incident neutron beam. The sample is heated above its melting point using two 125\,W CO$_{2}$ lasers pointing downwards onto the sample at an angle of 20\,$^{\circ}$ to the vertical to achieve a homogeneous temperature distribution. Temperature is recorded using an optical pyrometer and the sample is monitored from the top and the side using video cameras (inset images). Neutron diffraction patterns are measured by the D4c detector array consisting of 9 microstrip $^{3}$He gas detectors providing a very high counting rate stability. The entire detector bank (situated behind the sample belljar in the photograph) rotates to cover the $7^{\circ}$ gaps between detectors to provide an overall scattering angle accessibility range of $1.5^{\circ}\le2\theta\le137^{\circ}$.\label{Fig9-D4c}}
\end{figure*}
Although clear changes in relation to ordering on both intermediate- and short-range length scales are evident on vitrification of calcium aluminate liquids by SXRD, Ca-O correlations penetrate the first Al-O coordination shell, and overlap considerably with other atom-atom interactions at higher bond lengths, introducing uncertainty in determining the local aluminium and calcium coordination environments, particular at high temperatures \cite{Drewitt11,Hannon00,Mei08a,Mei08b}. It is, therefore, advantageous to apply element selective techniques to unambiguously measure the aluminium and calcium coordination environments. As discussed in the introduction, high-temperature liquid $^{27}$Al NMR spectroscopy measurements observe the fast exchange limit such that specific coordination environment populations cannot be resolved \cite{Cote92,Poe93,Poe94,Massiot95,Florian18}. NMR experiments using the quadrupolar spin-$\frac{7}{2}$ $^{43}$Ca nuclide are also limited by its low sensitivity and natural abundance \cite{Dupree97}, especially at high-temperature. Neutron diffraction with isotope substitution (NDIS) has previously been used to determine local calcium coordination environment in (CaO)$_{0.48}$(SiO$_{2}$)(Al$_{2}$O$_{3}$)$_{3}$ glass \cite{Eckersley88,Gaskell91}. NDIS requires good counting statistics and is thus limited by the sample size and available flux at neutron sources. Despite this limitation, NDIS has recently been successfully applied to small ($\sim$2-3\,mm diameter) liquid calcium aluminate \cite{Drewitt12b,Drewitt17} and silicate spherules  \cite{Skinner12}, notably including application of the ``double-difference'' method to directly measure the Ca-Ca atom-atom interactions in liquid C3A \cite{Drewitt17}.

The coherent scattering intensity measured by neutron diffraction is represented by the total structure factor
\begin{equation}\label{FQ-eq}
F(Q)=\sum_{\alpha}\sum_{\beta}c_{\alpha}c_{\beta}b_{\alpha}b_{\beta}[S_{\alpha\beta}(Q)-1],
\end{equation}
where $b$ denotes the coherent neutron scattering length. Natural calcium is composed mainly of $^{40}\rm{Ca}$ (96.941\,\% abundance) and $^{44}\rm{Ca}$ (2.086\,\% abundance) with coherent neutron scattering lengths of $4.80(2)$\,fm and $1.42(6)$\,fm, respectively \cite{Sears92}. If three structurally identical calcium-aluminate samples are prepared containing Ca in its natural isotopic abundance, predominantly $^{44}$Ca, and a 50:50 mixture of the two, then the $S_{\alpha\beta}(Q)$ functions involving Ca receive different weightings and give rise to observably different total structure factors denoted $^{\rm{nat}}F(Q)$, $^{\rm{mix}}F(Q)$, and $^{44}F(Q)$. The  $S_{\alpha\beta}(Q)$ with ${\alpha,\beta}\ne\rm{Ca}$ will, however, receive identical weighting in each $^{\rm{nat}}F(Q)$, $^{\rm{mix}}F(Q)$, or $^{44}F(Q)$. By linear combination of these $F(Q)$ it is possible to eliminate specific partial structure factors from the scattering function. For a calcium aluminate system, equation \ref{FQ-eq} can be represented by the pseudobinary combination
\begin{eqnarray}
\nonumber F(Q)&=&c^2_{Ca}b^2_{Ca}[S_{\rm{CaCa}}(Q)-1]\\
& &+2c_{Ca}b_{Ca}S_{Ca\mu}(Q)+S_{\mu\mu}(Q),\label{FQ-pseudo}
\end{eqnarray}
where
\begin{eqnarray}
\nonumber S_{\rm{Ca}\mu}(Q)&=&c_{\rm{Al}}b_{\rm{Al}}[S_{\rm{CaAl}}(Q)-1]\\
& &+c_{\rm{O}}b_{\rm{O}}[S_{\rm{CaO}}(Q)-1],\label{S_Camu}
\end{eqnarray}
and
\begin{eqnarray}
\nonumber S_{\mu\mu}(Q)&=&c_{\rm{Al}}^{2}b_{\rm{Al}}^{2}[S_{\rm{AlAl}}(Q)-1]+c_{\rm{O}}^{2}b_{\rm{O}}^{2}[S_{\rm{OO}}(Q)-1]\\
& &+2c_{\rm{Al}}c_{\rm{O}}b_{\rm{Al}}b_{\rm{O}}[S_{\rm{AlO}}(Q)-1].\label{S_mumu}
\end{eqnarray}
Expressed in matrix form,
\begin{eqnarray}\label{matrix}
\nonumber 
\left[\begin{array}{ccc}S_{\rm{CaCa}}(Q)-1\\
S_{\rm{Ca}\mu}(Q)\\
S_{\mu\mu}(Q)
\end{array}\right]&=&\left[\begin{array}{ccc}
c_{\rm{Ca}}^{2}b_{44}^{2} & 2c_{\rm{Ca}}b_{44} & 1\\
c_{\rm{Ca}}^{2}b_{\rm{mix}}^{2} & 2c_{\rm{Ca}}b_{\rm{mix}} & 1\\
c_{\rm{Ca}}^{2}b_{\rm{nat}}^{2} & 2c_{\rm{Ca}}b_{\rm{nat}} & 1\end{array}\right]^{-1}\\
& &\times
\left[\begin{array}{ccc}
^{44}F(Q)\\
^{\rm{mix}}F(Q)\\
^{\rm{nat}}F(Q)
\end{array}
\right].
\end{eqnarray}
It is, therefore, possible to separate the measured diffraction patterns into structure factors $S_{\rm{Ca}\mu}(Q)$, which contains all correlations involving Ca apart from with itself, $S_{\mu\mu}(Q)$ in which all Ca correlations have been eliminated, and a direct measurement of the $S_{\rm{CaCa}}(Q)$ partial structure factor.  The corresponding real-space functions $g_{\rm{CaCa}}(r)$, $g_{\rm{Ca}\mu}(r)$, and $g_{\mu\mu}(r)$, are obtained using the Fourier transform relation defined by equation \ref{Gr-FT}. Mean coordination numbers are obtained by integrating over a peak in real-space arising from specific atom-atom pairs of interest \cite{Fischer06}.

\begin{figure}
\centering
\includegraphics[width=0.48\textwidth]{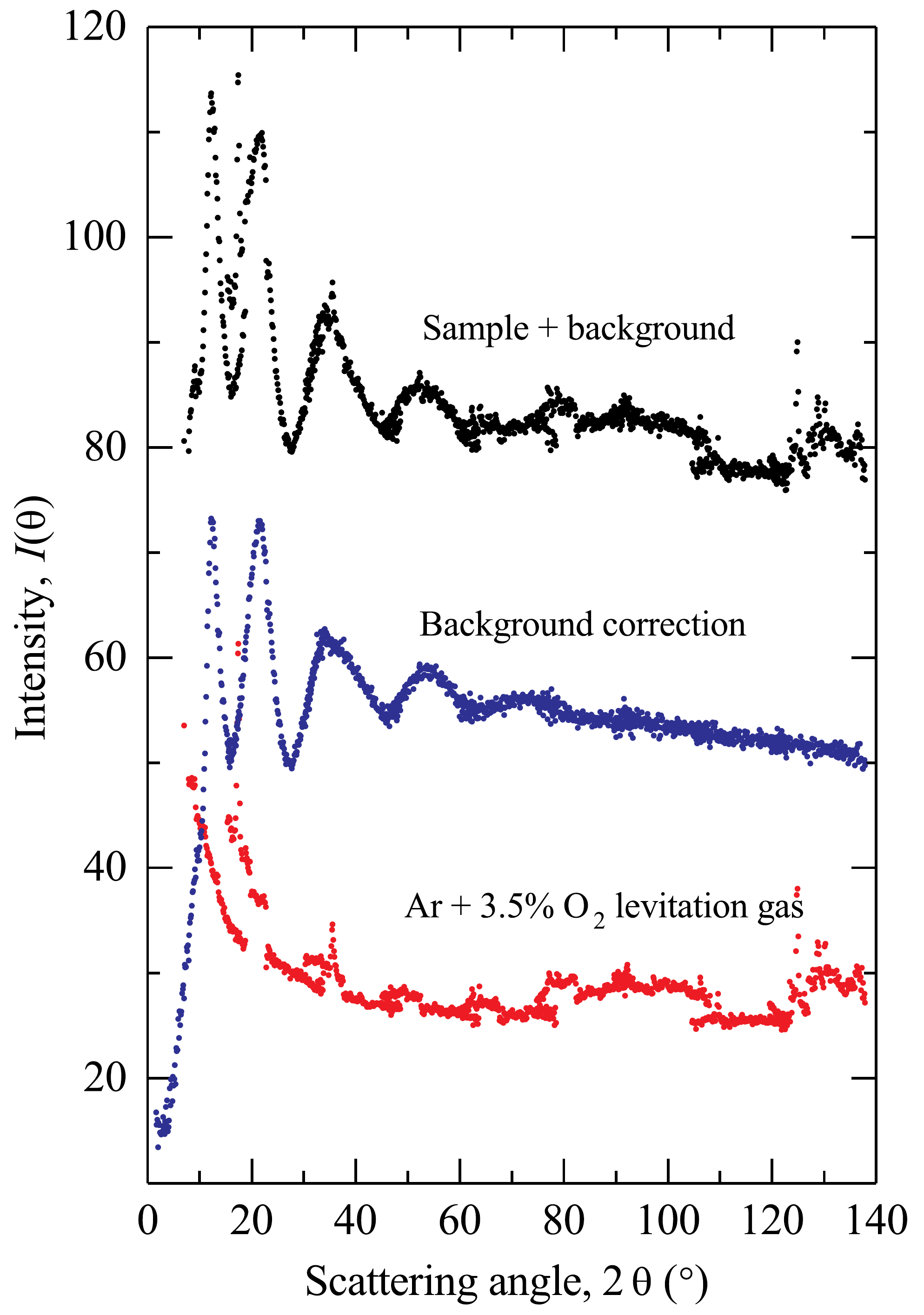}
\caption{The measured neutron diffraction patterns for levitated liquid CaAl$_{2}$O$_{4}$ (CA) at 1700$^{\circ}$C, the empty levitation device inside the diffraction chamber with a flow of $96.5$\,\%\,Ar$+3.5$\,\%\,O$_{2}$ gas, and the background corrected signal.\label{Fig10-NDIS-RAW}}
\end{figure}
Aerodynamic levitation with laser heating was combined with $^{44}$Ca NDIS to measure the structure of levitated CA \cite{Drewitt12b} and C3A \cite{Drewitt17} liquids and glasses, using the experimental set-up shown in figure \ref{Fig9-D4c} \cite{Hennet06} installed at the D4c neutron diffractometer \cite{Fischer02} at the Institut Laue-Langevin (ILL), France. The ILL delivers a very uniform flux of neutrons with high-count rates. As a result, D4c is exceptionally stable and capable of detecting small changes in scattering between samples. However, the small size of the levitated samples (2-3\,mm diameter) is considerably less than typical D4c sample dimensions of 7\,mm diameter and 50\,mm height \cite{Fischer02}. Long counting times of up to $24$\,h were therefore required per sample, with up to $40$\,h for the mixed isotope sample due to its double weighting in the difference function. The liquid spherules remained stably levitated with no observable mass loss or change in scattering intensity outside of statistical error during the course of the experiment. The glasses were measured under vacuum with the sample resting on top of the levitation nozzle. The as-measured diffraction patterns for levitated liquid CA and the empty levitation device inside the diffraction chamber with a flow of $96.5$\,\%\,Ar$+3.5$\,\%\,O$_{2}$ gas are shown in figure \ref{Fig10-NDIS-RAW}. Since the nozzle was completely hidden by neutron-absorbing B$_{4}$C collimation flags, and with no signal from a sample container to correct for, very clean neutron diffraction patterns are obtained with the relatively low background intensity originating predominantly from gas scattering in the chamber from the levitating gas flow. 

\begin{figure}
\centering
\includegraphics[width=0.48\textwidth]{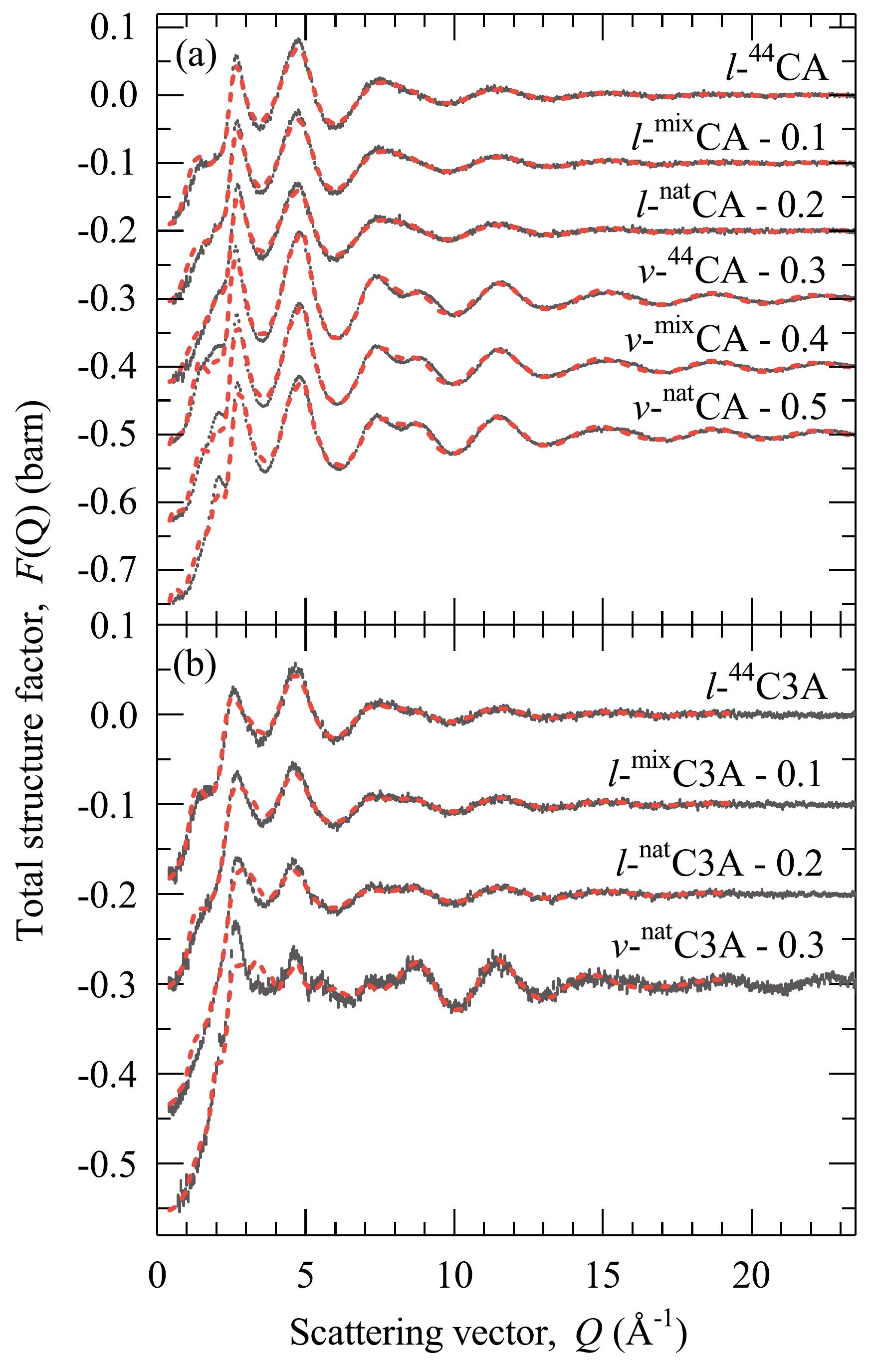}
\caption{Total structure factors $^{44}F(Q)$, $^{\rm{mix}}F(Q)$, and $^{\rm{nat}}F(Q)$ obtained by neutron diffraction for the (CaO)$_{x}$(Al$_{2}$O$_{3}$)$_{1-x}$ liquids (a) $x=0.5$ (CA) at $1700^{\circ}$C \cite{Drewitt12b} and (b) $x=0.75$ (C3A) at $1800^{\circ}$C \cite{Drewitt17} denoted by the prefix $l$-, together with the room temperature glass (vitreous) samples denoted by the prefix $v$- (vertical error bars). The red dashed curves are the $F(Q)$ functions generated from the AIM-MD simulations obtained for CA and C3A liquids at $2230^{\circ}$C \cite{Drewitt11} and CA \cite{Drewitt12b} and C3A glasses at $30^{\circ}$C.\label{Fig11-FQ-NDIS}}
\end{figure}
\begin{figure}
\centering
\includegraphics[width=0.48\textwidth]{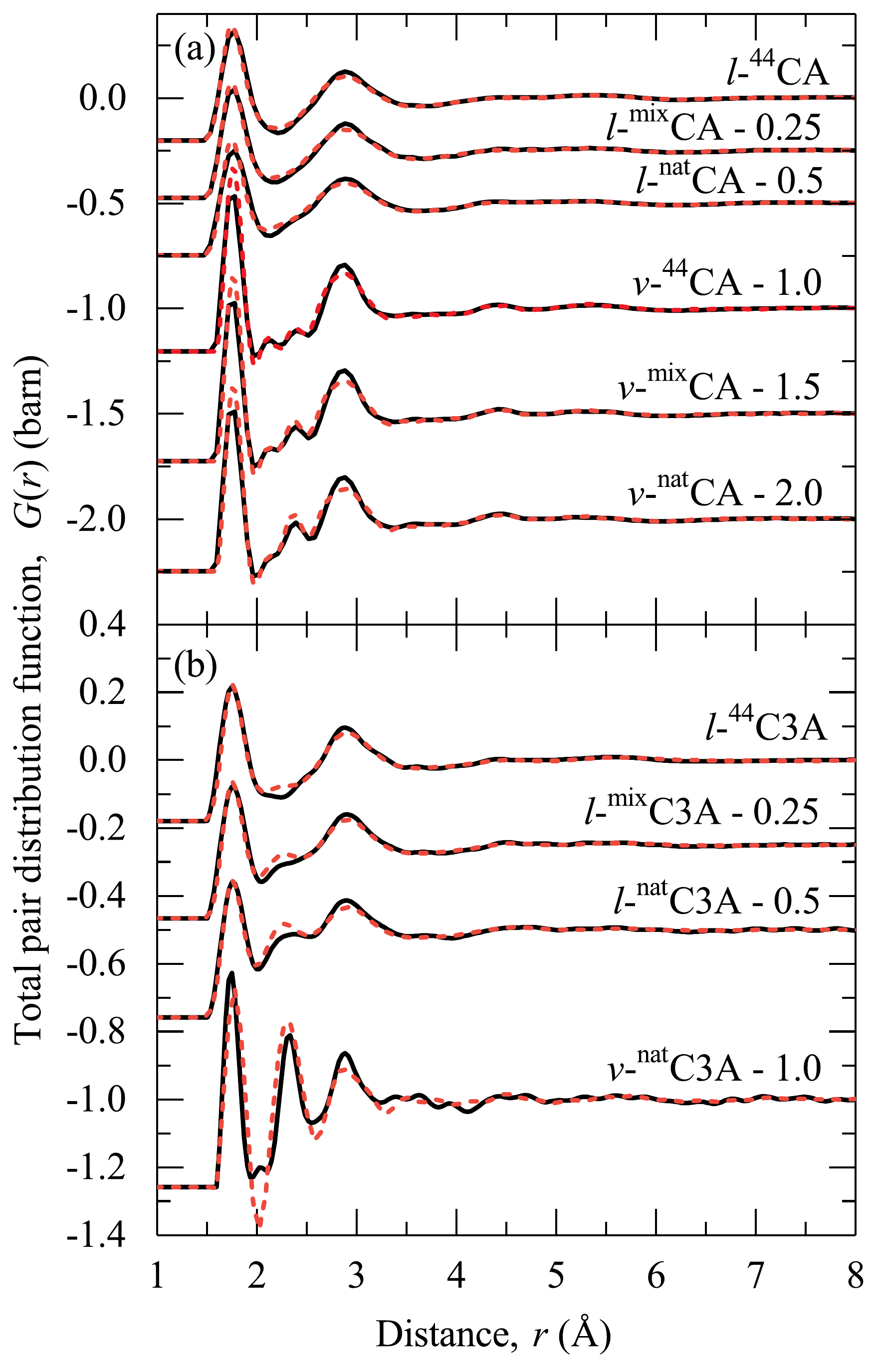}
\caption{Total pair distribution functions $^{44}G(r)$, $^{\rm{mix}}G(r)$, and $^{\rm{nat}}G(r)$ for the (CaO)$_{x}$(Al$_{2}$O$_{3}$)$_{1-x}$ liquids (a) $x=0.5$ (CA) and (b) $x=0.75$ (C3A) denoted by the prefix $l$-, together with the room temperature glass (vitreous) samples denoted by the prefix $v$-, as obtained by Fourier transforming the the corresponding $F(Q)$ functions in figure \ref{Fig11-FQ-NDIS} from neutron diffraction measurements (solid black curves) and AIM-MD simulations (red dashed curves).\label{Fig12-Gr-NDIS}}
\end{figure}
The total neutron structure factors $^{\rm{nat}}F(Q)$, $^{\rm{mix}}F(Q)$, and $^{44}F(Q)$ measured for the isotopically substituted CA and C3A liquids and glasses are shown in figure \ref{Fig11-FQ-NDIS}. The peak arising from cation-cation correlations at $\sim2.2\,\rm{\AA}^{-1}$ in the SXRD measurements (figure \ref{Fig5-SXRD-SQ}) is absent in the neutron $F(Q)$ functions due to the lower neutron-scattering cross sections for the cations compared to x-ray scattering.  The corresponding $^{\rm{nat}}G(r)$, $^{\rm{mix}}G(r)$, and $^{44}G(r)$ functions are shown in figure \ref{Fig12-Gr-NDIS}. All real-space peak positions and average coordination numbers are listed in table \ref{Table2}. The first peak in the $G(r)$ functions arises from the nearest neighbour Al-O correlations. The main difference between the $G(r)$ functions for CA and C3A is the second peak arising from Ca-O correlations at $\sim2.3\,\rm{\AA}$ is much stronger in the C3A measurements due to the higher fraction of CaO. This peak is strongest in the glass measurements and is barely discernible in the liquid CA measurements. The Ca-O peak progressively weakens from $^{\rm{nat}}G(r)$ to $^{\rm{mix}}G(r)$ and $^{44}G(r)$ due to the reducing calcium neutron scattering length. 

\begin{figure}
\centering
\includegraphics[width=0.48\textwidth]{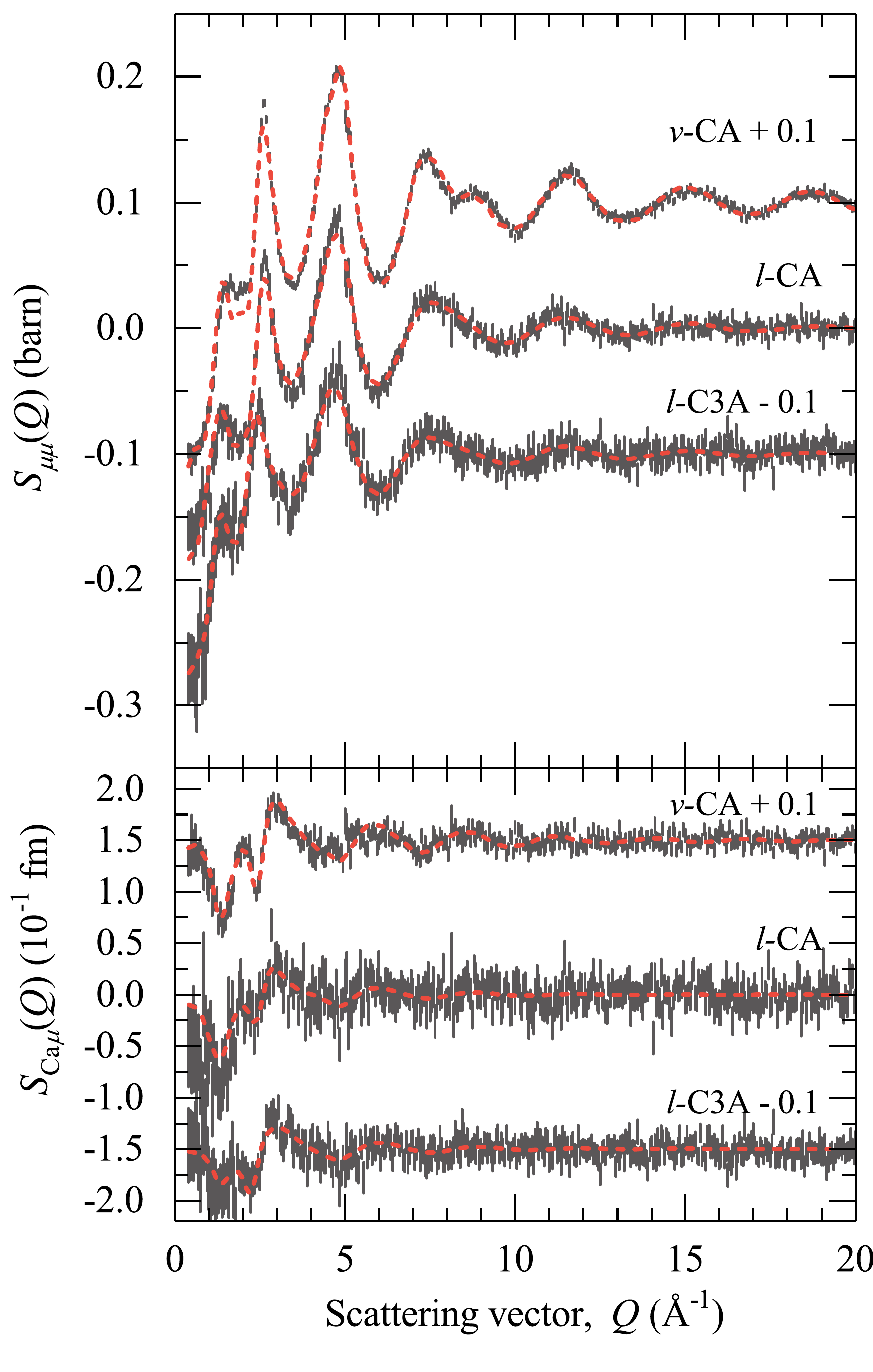}
\caption{The reciprocal-space structure factors $S_{\mu\mu}(Q)$ and $S_{\rm{Ca}\mu}(Q)$ for the CA \cite{Drewitt12b} and C3A \cite{Drewitt17} liquids (denoted by the prefix $l$-) and CA glass \cite{Drewitt12b} (denoted by the prefix $v$-), as calculated by a linear combination of the $F(Q)$ functions shown in figure \ref{Fig11-FQ-NDIS} (data points with error bars). The red dashed curves are the functions calculated from the AIM-MD simulations \cite{Drewitt11,Drewitt12b}.\label{Fig13-NDIS-DiffsQ}}
\end{figure}
\begin{figure}
\centering
\includegraphics[width=0.48\textwidth]{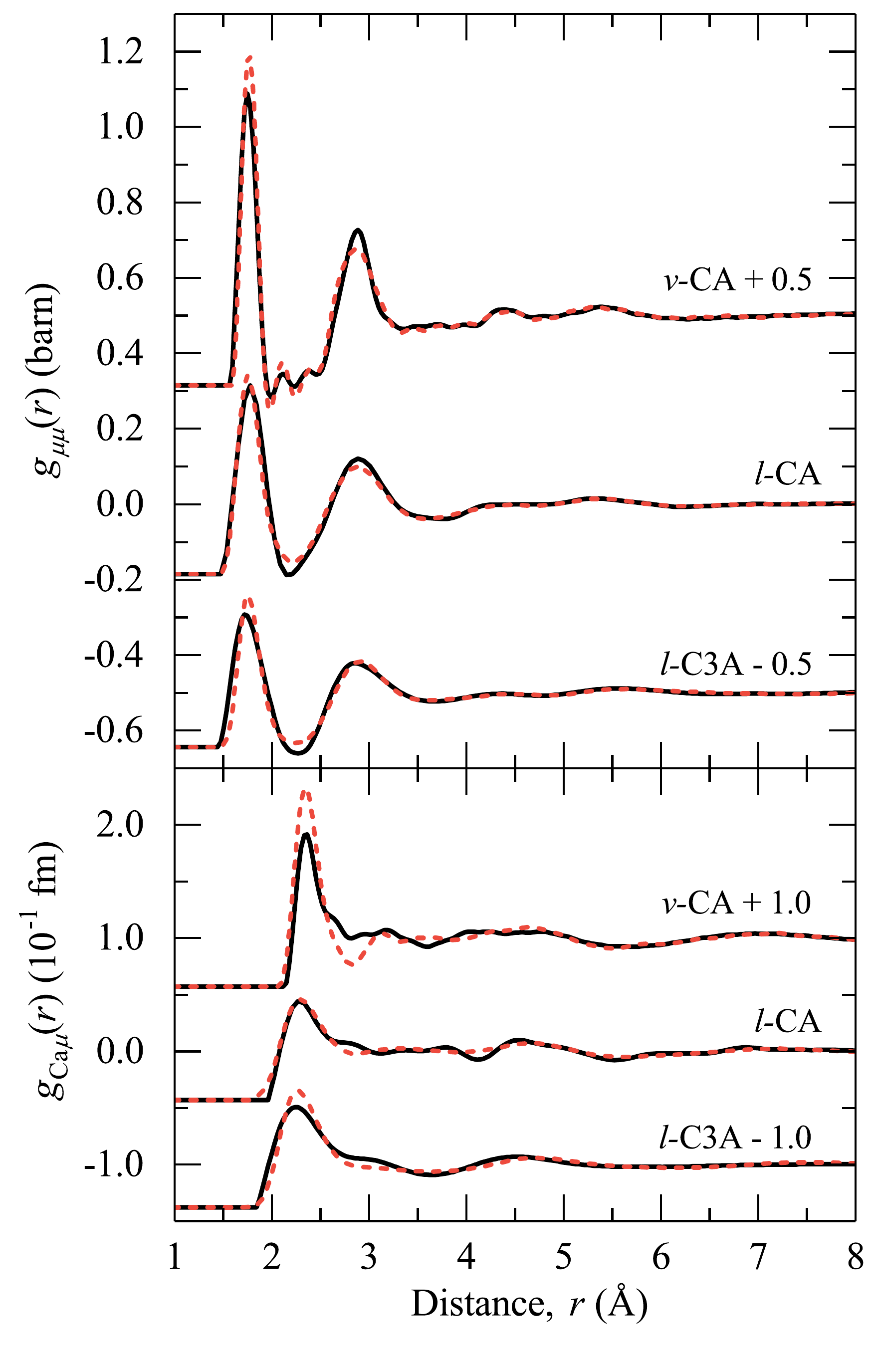}
\caption{The real-space pair distribution function $g_{\mu\mu}(r)$ and $g_{\rm{Ca}\mu}(r)$ for the CA \cite{Drewitt12b} and C3A \cite{Drewitt17} liquids (denoted by the prefix $l$-) and CA glass \cite{Drewitt12b} (denoted by the prefix $v$-) as obtained by Fourier transforming the experimental \cite{Drewitt12b,Drewitt17} (solid black curves) and AIM-MD \cite{Drewitt11,Drewitt12b} (dashed red curves) difference functions shown in figure \ref{Fig13-NDIS-DiffsQ}.\label{Fig14-NDIS-Diffsr}}
\end{figure}
\begin{figure}
\centering
\includegraphics[width=0.48\textwidth]{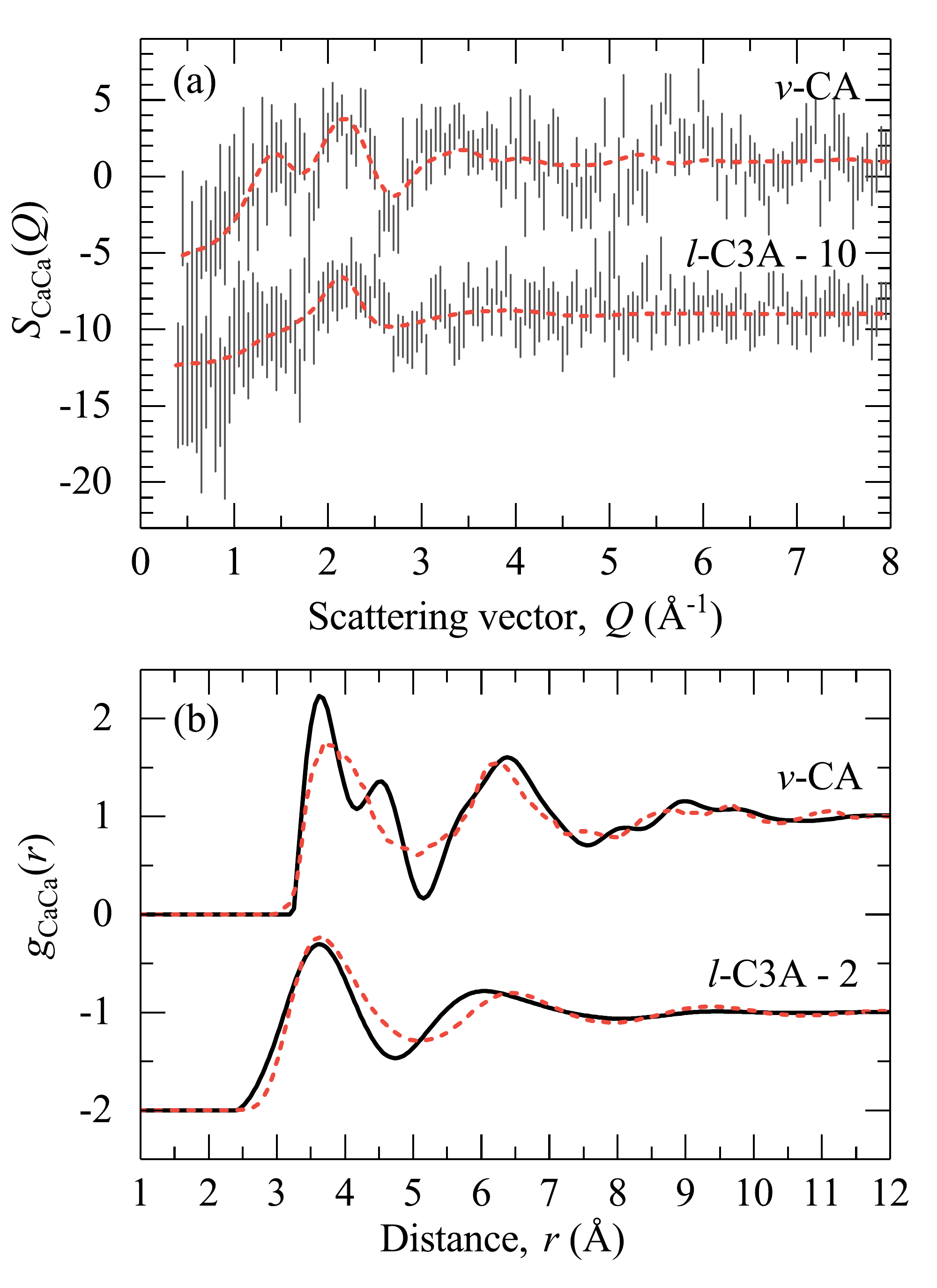}
\caption{(a) The partial structure factor $S_{\rm{CaCa}}(Q)$  (data points with error bars) and (b) partial pair distribution function $g_{\rm{CaCa}}(r)$ (solid black curves) determined from NDIS experiments for glassy (vitreous) $v$-CA \cite{Drewitt12b} and liquid $l$-C3A \cite{Drewitt17}. The dashed red curves show the results generated directly from AIM-MD simulations \cite{Drewitt11,Drewitt12b}.
\label{Fig15-CaCa}}
\end{figure}
The reciprocal-space structure factors $S_{\mu\mu}(Q)$ and $S_{\rm{Ca}\mu}(Q)$ obtained by linear combination of $^{\rm{nat}}F(Q)$, $^{\rm{mix}}F(Q)$, and $^{44}F(Q)$ according to equation \ref{matrix}, are shown in figure \ref{Fig13-NDIS-DiffsQ} and the corresponding real-space functions $g_{\mu\mu}(r)$ and $g_{\rm{Ca}\mu}(r)$ are shown in figure \ref{Fig14-NDIS-Diffsr}. The $g_{\mu\mu}(r)$ functions contain contributions arising solely from $\mu$-$\mu$ ($\mu=\rm{Al},\rm{O})$ correlations and have a first peak arising from Al-O nearest-neighbours. The $g_{\rm{Ca}\mu}(r)$ functions contain contributions arising from Ca-Ca and Ca-$\mu$ correlations only and their first peak corresponds to Ca-O nearest-neighbours. The NDIS measurements for CA glass \cite{Drewitt12b} and liquid C3A \cite{Drewitt17} were of suitable quality to enable the direct extraction of the Ca-Ca partial structure factor $S_{\rm{CaCa}}(Q)$ and corresponding partial-pair correlation function $g_{\rm{CaCa}}(r)$ which are shown in figure \ref{Fig15-CaCa}.

Overall, the AIM-MD simulations are in excellent agreement with the experimental data. However, in the $F(Q)$ and $G(r)$ functions for C3A, there are some regions of noticeable discrepancies, particularly in the low-$Q$ region in reciprocal space and in the region of the Ca-O peak in real-space. To evaluate these small observed discrepancies, the structural model obtained from AIM-MD for liquid C3A was refined by RMC methods including data from the NDIS experiments and SXRD \cite{Drewitt11,Drewitt17}. The RMC refinement made only subtle changes to the AIM-MD-derived configuration to achieve a better fit to the experimental results, and the basic structure including coordination numbers and bond angle distributions was largely unchanged \cite{Drewitt17}, confirming the reliability of the AIM-MD atomistic model.

\section{Discussion}

The combined strengths of SXRD and NDIS experiments and AIM-MD simulations reported and reviewed here, provide a detailed overview of the structural processes that take place in these multicomponent aluminate liquids during vitrification. Overall the AIM-MD results are in excellent agreement with the experimental findings. Examination of the AIM-MD trajectories reveals the structure of liquid CA contains 12\,\% non-bridging oxygens, with a significant fraction of AlO$_{5}$ units (15\,\%) to maintain local charge balance accompanied by small concentrations of 3- and 6-fold coordinations ($\le\,2$ and 0.4\,\%, respectively), while the remaining aluminium forming AlO$_{4}$ tetrahedral units. Considering all Al-O pairs, 18\,\% of oxygen atoms are coordinated by more than two aluminium atoms with 7(1)\,\% involving formal triclusters in which one oxygen atom is shared by three aluminium tetrahedra \cite{Lacy63}. Calcium has a broad distribution of coordination sites from 4- to 9-fold by oxygen, with an average coordination number $\bar{n}_{\rm{Ca}}^{\rm{O}}=6.2$, forming small clusters of edge- and face-sharing Ca-centred polyhedra. On vitrification, the structure of CA glass is characterised by a predominantly corner-shared tetrahedral network. However, the AIM-MD model predicts 7\,\% of all oxygen atoms in the glass are non-bridging with $\sim4$\,\% AlO$_{5}$ units and 5\,\% formal triclusters (12\,\% considering all Al-O pairs), consistent with 5\,\% triclusters detected in the glass by heteoronuclear correlation NMR spectroscopy \cite{Iuga05}. Edge- and face-sharing Ca-centred polyhedra, with an average coordination number $\bar{n}_{\rm{Ca}}^{\rm{O}}=6.2$, form large branched chains that weave through the glass network contributing to cationic ordering on an intermediate range length-scale. On cooling the liquid, the AIM-MD simulations indicate that maximum cluster size increases significantly in the vicinity of the dynamical crossover temperature between $\sim1250\,^{\circ}$C and $1000\,^{\circ}$C, coinciding with inflections observed in the cooling curve at around $1.25T_{\rm{g}}$ and changes in the first peak parameters in $S(Q)$ indicative of a development of ordering between cation-centred polyhedra. We note, however, that the simulated glass structure is sensitive to the quench rate. The cooling rate of $10^{12}$\,K\,s$^{-1}$ employed in the AIM-MD simulations is ten orders of magnitude higher than the experimental value such that the fictive temperature is likely to be much higher in the simulations. There are some slight indications of disagreement between the NDIS measurements and AIM-MD model of CA glass, such as in the AIM-MD derived $g_{\rm{CaCa}}$ which exhibits a single peak at $r_{\rm{CaCa}}\simeq3.78\,\rm{\AA}$, compared to two clearly resolved peaks measured by NDIS at $r_{\rm{CaCa}}=3.59$ and 4.41\,$\rm{\AA}$. Also, the height of the first peak in the AIM-MD derived $S(Q)$ functions for both CA and C3A glass are slightly lower than in the SXRD measurements, indicating a more liquid-like degree of ordering in the simulated glass structures.

The results show the structure of liquid C3A is largely composed of AlO$_{4}$ tetrahedra (93\,\%). The C3A composition has an O/Al ratio of 3, such that in a network of AlO$_{4}$ there should be a mean number of 2 bridging oxygen atoms per aluminium and the overall fraction of non-bridging oxygen atoms is 2/3 with 1/3 bridging oxygens \cite{Skinner12b}. We find 60\,\% non-bridging oxygen and 36\,\% bridging oxygen atoms indicating a slightly higher fraction of bridging oxygens expected from a simple network model and consistent with a large fraction of non-bridging oxygens detected by x-ray absorption and NMR spectroscopy methods \cite{Neuville08,Neuville10}. This difference is accounted for by the presence of 3-4\,\% 'free oxygen' ions, which are not bonded to aluminium, and $\sim$1\,\% oxygen atoms existing in triclusters. Although the majority of AlO$_{4}$ tetrahedra belong to a single infinitely connected major corner-shared cluster, around 15 to 20\,\% aluminium are connected to smaller clusters, with around 10\,\% forming Al$_{2}$O$_{7}$ dimers or isolated AlO$_{4}$ tetrahedral units. Calcium has a wide distribution of coordination sites from 4- to 8-fold by oxygen, with a slightly smaller an average coordination number of $\bar{n}_{\rm{Ca}}^{\rm{O}}=5.6$ than for liquid CA. All CaO polyhedra are connected by corners to a single network with 90\,\% edge- and face-sharing connectivity. Highly coordinated CaO$_{y}$ polyhedra  ($y=6,7,8$) preferentially bond to a higher fraction of bridging oxygen neighbours and have higher coordination by aluminium compared to smaller CaO$_{y}$ polyhedra ($y=4,5$). The simulated $g_{\rm{CaCa}}(r)$ is in very good agreement with the NDIS measurement, giving an average coordination number $\bar{n}_{\rm{Ca}}^{\rm{Ca}}=8.5$. The C3A glass structure obtained in the AIM-MD simulations is characterised by 99\,\% of all Al being tetrahedrally coordinated and very few AlO$_5$ units. Compared to the melt the connectivity of AlO$_4$ tetrahedra increases to 85\,\% of the tetrahedra being part of a large cluster and only 5\,\% of isolated tetrahedra or dimers. The number of 'free oxygens' is reduced to about 2\,\% and OAl$_3$ triclusters are virtually absent. $g_{CaCa}(r)$ of the simulated C3A glass shows at least two contributions to the first peak with a maximum at 3.48\,{\AA} and a strong shoulder at 3.80\,{\AA}. These distances represent the characteristic Ca-Ca distances between edge- and corner-sharing Ca polyhedra, respectively. 

\section{Conclusions}

Although liquid CA (CaAl$_{2}$O$_{4}$) defies Zachariasen's rules for glass formation, the liquid structure reorganises on quenching to form a predominantly corner-shared network structure based on AlO$_{4}$ tetrahedra via the breakdown of AlO$_{5}$ and oygen triclusters in the liquid. This reorganisation is accompanied by changes in medium range order via formation of edge- and face-shared chains of Ca-centred polyhedra. At the Al$_{2}$O$_{3}$-rich end of the glass forming region, there is an increase in the concentration of AlO$_{5}$ polyhedra in the liquid structure, while the lifetime of AlO$_{4}$ tetrahedra is small due to fast oxygen hopping between four- and five-fold coordinated Al \cite{Drewitt11}.

At the CaO-rich end of the glass forming region, liquid C3A (Ca$_{3}$Al$_{2}$O$_{6}$) although significantly de-polymerised is still largely composed of AlO$_{4}$ tetrahedra, most of which belong to an infinite corner-shared network. The results indicate the presence of about 10\,\% unconnected Al$_{2}$O$_{7}$ and AlO$_{4}$ monomers and dimers in the liquid. The number of these isolated units is expected to increase with CaO concentration, such that the upper value of the glass-forming composition could be described in terms of a percolation threshold at which the glass can no longer support the formation of an infinitely connected AlO$_{4}$ network.

Overall, the AIM-MD simulations are in excellent agreement with the SXRD and NDIS experiments suggesting an accurate potential model. However, small discrepancies between in the finer structural details between the simulated glasses and experimental measurements are apparent indicating a small degree of liquid-like ordering persists in the simulated glass trajectories. This may be due to the short simulation time-scales which are unrepresentative of the viscous kinetic processes involved in supercooling and glass formation.  This limitation could be overcome by future advances in simulation methods, for example by integrating rare event sampling techniques based on large deviation theory \cite{Turci17} into MD simulation codes to explore the energy landscape in the supercooled region. For the aluminate liquids studied here, this could involve sampling trajectories that preferentially explore a more structured glass $g_{\rm{CaCa}}(r)$, as measured by NDIS compared to the AIM-MD computed function \cite{Drewitt12b}, thereby accelerating the sampling of configurations representative of the glassy state.

\ack
We thank Aleksei Bytchkov (ESRF) for his assistance in the SXRD data collection at beamline ID11 (ESRF beamtime award HD387), Henry Fischer (ILL) for his help and expert assistance in the neutron diffraction data collection, and Dominique Thiaudi\`{e}re for collection of the SXRD data for C3A glass at the Diffabs beamline at the Soleil Synchrotron. We also thank Daniel Neuville (IPGP-Paris) and Adrian Barnes (Bristol) for helpful discussions and the provision of starting materials, we gratefully acknowledge the contributions of Anita Zeidler (Bath) and Simon Kohn (Bristol) to the NDIS data collection, and thank Francesco Turci and Paddy Royall for helpful comments. The SXRD experiments were made in the framework of the French National Research Agency (ANR) grant no. NT09 432740 awarded to LH. JD is supported financially by NERC standard grant NE/P002951/1.
\section*{References}
\bibliography{bibliography}

\begin{thebibliography}{10}

\bibitem{McMillan96}
P.~F. McMillan, W.~T. Petuskey, B.~Cot\'{e}, D.~Massiot, C.~Landron, and J.~P.
  Coutures.
\newblock {\em J. Non-Cryst. Solids}, 195:261, 1996.

\bibitem{Shelby89}
J.~E. Shelby, C.~M. Shaw, and M.~S. Spess.
\newblock {\em J. Appl. Phys.}, 66:1149, 1989.

\bibitem{Weber98}
J.~K.~R. Weber, J.~J. Felten, B.~Cho, and P.~C. Nordine.
\newblock {\em Nature}, 393:769, 1998.

\bibitem{Aizawa04}
H.~Aizawa, H.~Uchiyama, T.~Katsumata, S.~Komuro, T.~Morikawa, H.~Ishizawa, and
  E.~Toba.
\newblock {\em Meas. Sci. Technol.}, 15:1484, 2004.

\bibitem{Haladejova16}
K.~Haladejov\'{a}, A.~Prnov\'{a}, R.~Klement, W.~H. Tuan, S.~J. Shih, and
  D.~Galusek.
\newblock {\em J. Eur. Ceram. Soc.}, 36:2969, 2016.

\bibitem{Eeckhout10}
K.~V. den Eeckhout, P.~F. Smet, and D.~Poelman.
\newblock {\em Materials}, 3:2536, 2010.

\bibitem{Zachariasen32}
W.~H. Zachariasen.
\newblock {\em J. Am. Chem. Soc.}, 54:3841, 1932.

\bibitem{Skinner13a}
L.B. Skinner, A.~C. Barnes, P.~S. Salmon, L.~Hennet, H.~E. Fischer, C.~J.
  Benmore, S.~Kohara, R.~J.~K. Weber, A.~Bytchkov, M.~C. Wilding, J.~B. Parise,
  T.~O. Farmer, I.~Pozdnyakova, S.~K. Tumber, and K.~Ohara.
\newblock {\em Phys. Rev. B}, 87:024201, 2013.

\bibitem{Nurse65}
R.~W. Nurse, J.~H. Welch, and A.~J. Majumdar.
\newblock {\em Trans. Br. Ceram. Soc.}, 64:409, 1965.

\bibitem{Massiot98}
D.~Massiot, B.~Touzo, D.~Trumeau, I.~Farnan, J.~C. Rifflet, C.~Bessada,
  A.~Douy, and J.~P. Coutures.
\newblock Time {R}esolved {V}ery {H}igh {T}emperature {NMR} {S}tudy of the
  {C}ooling {P}rocess of {C}a{O}-{A}l$_{2}${O}$_{3}$ {L}iquids.
\newblock In P.~Colombet, A.~R. Grimmer, H.~Zanni, and P.~Sozzani, editors,
  {\em Nuclear {M}agnetic {R}esonance {S}pectroscopy of {C}ement-{B}ased
  {M}aterials}, pages 107--116. Springer Verlag Berlin, 1998.

\bibitem{McMillan83}
P.~McMillan and B.~Piroou.
\newblock {\em J. Non-Cryst. Solids}, 55:221, 1983.

\bibitem{Angell95}
C.~A. Angell.
\newblock {\em Science}, 267:1924, 1995.

\bibitem{Urbain83}
G.~Urbain.
\newblock {\em Rev. Int. Hautes Temp. Refract. Fr.}, 20:135, 1983.

\bibitem{Drewitt12b}
J.~W.~E. Drewitt, L.~Hennet, A.~Zeidler, S.~Jahn, P.~S. Salmon, D.~R. Neuville,
  and H.~E. Fischer.
\newblock {\em Phys. Rev. Lett}, 109:235501, 2012.

\bibitem{Bohmer93}
R.~Bohmer, K.~L. Ngai, C.~A. Angell, and D.~J. Plazek.
\newblock {\em J. Chem. Phys}, 99:4201.

\bibitem{Vogel21}
H.~Vogel.
\newblock {\em Phys. Z}, 22:645, 1921.

\bibitem{Fulcher25}
G.~S. Fulcher.
\newblock {\em J. Amer. Ceram. Soc.}, 8:339, 1925.

\bibitem{Tammann26}
G.~Tammann and W.~Hesse.
\newblock {\em Z. Anorg. Allg. Chem.}, 156:245, 1926.

\bibitem{Angel85}
C.~A. Angel.
\newblock {\em J. Non-Cryst. Solids}, 73:1, 1985.

\bibitem{Debenedetti01}
P.~B. Debenedetti and F.~H. Stillinger.
\newblock {\em Nature}, 420:259, 2001.

\bibitem{Greaves07}
G.~N. Greaves and Sen. S.
\newblock {\em Adv. Phys.}, 56:1, 2007.

\bibitem{Gotze92}
W.~Gotze and L.~Sjogren.
\newblock {\em Rep. Prog. Phys.}, 55:241, 1992.

\bibitem{Anderson05}
H.~C. Anderson.
\newblock {\em Proc. Nat. Acad. Sci}, 102:6686, 2005.

\bibitem{Ossi06}
P.~M. Ossi.
\newblock {\em Disordered materials: an introduction}.
\newblock Berlin; New York: Springer, 2nd edition, 2006.

\bibitem{Goldstein69}
M.~Goldstein.
\newblock {\em J. Chem. Phys.}, 51:3728, 1969.

\bibitem{Hannon00}
A.~C. Hannon and J.~M. Parker.
\newblock {\em J. Non-Cryst. Solids}, 274:102, 2000.

\bibitem{Benmore03}
C.~J. Benmore, J.~K.~R. Weber, S.~Sampath, J.~Siewenie, J.~Urquidi, and J.~A.
  Tangeman.
\newblock {\em J. Phys.: Condens. Matter}, 15:S2413, 2003.

\bibitem{Poe93}
B.~T. Poe, P.~F. McMillan, B.~Cot\'{e}, D.~Massiot, and J.~P. Coutures.
\newblock {\em Science}, 259:786, 1993.

\bibitem{Poe94}
B.~T. Poe, P.~F. McMillan, B.~Cot\'{e}, D.~Massiot, and J.~P. Coutures.
\newblock {\em J. Am. Ceram. Soc.}, 77:1832, 1994.

\bibitem{Massiot95}
D.~Massiot, D.~Trumeau, B.~Touzo, I.~Farnan, J.~C. Rifflet, A.~Douy, and J.~P.
  Coutures.
\newblock {\em J. Phys. Chem.}, 99:16455, 1995.

\bibitem{Weber03}
J.~K.~R. Weber, C.~J. Benmore, J.~A. Tangeman, J.~Siwenie, and K.~J. Hiera.
\newblock {\em J. Neutron. Res}, 11:113, 2003.

\bibitem{Hennet07a}
L.~Hennet, I.~Pozdnyakova, V.~Cristiglio, G.~J. Cuello, S.~Jahn, S.~Krishnan,
  M.~L. Saboungi, and D.~L. Price.
\newblock 19:455210, J. Phys.: Condens. Matter.

\bibitem{Mei08b}
Q.~Mei, C.~J. Benmore, J.~K.~R. Weber, M.~Wilding, J.~Kim, and J.~Rix.
\newblock {\em J. Phys.: Condens. Matter}, 20:245107, 2008.

\bibitem{Cristiglio10}
V.~Cristiglio, L.~Hennet, G.~J. Cuello, I.~Pozdnyakova, M.~R. Johnson, H.~E.
  Fischer, D.~Zanghi, and D.~L. Price.
\newblock {\em J. Non-Cryst. Solids}, 356:2492, 2010.

\bibitem{Drewitt11}
J.~W.~E. Drewitt, S.~Jahn, V.~Cristiglio, A.~Bytchkov, M.~Leydier,
  S.~Brassamin, H.~E. Fischer, and L.~Hennet.
\newblock {\em J. Phys.: Condens. Matter}, 23:155101, 2011.

\bibitem{Drewitt12a}
J.~W.~E. Drewitt, S.~Jahn, V.~Cristiglio, A.~Bytchkov, M.~Leydier,
  S.~Brassamin, H.~E. Fischer, and L.~Hennet.
\newblock {\em J. Phys.: Condens. Matter}, 24:099501, 2012.

\bibitem{Drewitt17}
J.~W.~E. Drewitt, A.~C. Barnes, S.~Jahn, S.~C. Kohn, M.~J. Walter, A.~N.
  Novikov, D.~R. Neuville, H.~E. Fischer, and L.~Hennet.
\newblock {\em Phys. Rev. B}, 95:064203, 2017.

\bibitem{Brandt01}
E.~H Brandt.
\newblock {\em Nature}, 413:474, 2001.

\bibitem{Marzo17}
A.~Marzo, A.~C. Barnes, and B.~W. Drinkwater.
\newblock {\em Rev. Sci. Instrum.}, 88:085105, 2017.

\bibitem{Krishnan97}
S.~Krishnan, J.~J. Felton, J.~E. Rix, J.~K.~R. Weber, P.~C. Nordine, M.~A.
  Beno, S.~Ansell, and D.~L. Price.
\newblock {\em Rev. Sci. Instrum.}, 68:3512, 1997.

\bibitem{Jacobs96}
G.~Jacobs, I.~Egry, K.~Maier, D.~Platzek, J.~Reske, and R.~Frahm.
\newblock {\em Rev. Sci. Instrum.}, 67:3683.

\bibitem{Paradis08}
P.~F. Paradis, T.~Ishikawa, and S.~Yoda.
\newblock {\em Adv. Space. Res.}, 41:2118.

\bibitem{Neuman04}
K.~C. Neuman and S.~M Block.
\newblock {\em Rev. Sci. Instrum.}, 75:2787, 2004.

\bibitem{Price10}
D.~L. Price.
\newblock {\em High-Temperature Levitated Materials}.
\newblock Cambridge University Press, 2010.

\bibitem{Hennet11a}
L.~Hennet, V.~Cristiglio, J.~Kozaily, I.~Pozdnyakova, H.~E. Fischer,
  A.~Bytchkov, J.~W.~E. Drewitt, M.~Leydier, D.~Thiaudi\`{e}re, S.~Gruner,
  S.~Brassamin, D.~Zanghi, G.~J. Cuello, M.~Koza, S.~Magaz\`{u}, G.~N. Greaves,
  and D.~L. Price.
\newblock {\em Eur. Phys. J. Special Topics}, 196:151, 2011.

\bibitem{Benmore17}
C.~J. Benmore and J.~K.~R. Weber.
\newblock {\em Adv. Phys. X}, 2:717, 2017.

\bibitem{Jahn07}
S.~Jahn and P.~M. Madden.
\newblock {\em Phys. Earth Planet. Int.}, 162:129, 2007.

\bibitem{Tang84}
K.~T. Tang and J.~P. Toennies.
\newblock {\em J. Chem. Phys.}, 80:3726, 1984.

\bibitem{Wilson96}
M.~Wilson, P.~A. Madden, and B.~J. Costa-Cabral.
\newblock {\em J. Phys. Chem.}, 100:1227, 1996.

\bibitem{Stone96}
A.~J. Stone.
\newblock {\em The theory of intermolecular forces}.
\newblock Oxford University Press, 1996.

\bibitem{Jahn08}
S.~Jahn.
\newblock {\em Am. Mineral.}, 93:1486, 2008.

\bibitem{Adjaoud08}
O.~Adjaoud, G.~Steinle-Neumann, and S.~Jahn.
\newblock {\em Chem. Geol.}, 256:184, 2008.

\bibitem{Drewitt15}
J.~W.~E. Drewitt, S.~Jahn, C.~Sanloup, C.~de~Grouchy, G.~Garbarino, and
  L.~Hennet.
\newblock {\em J. Phys.: Condens. Matter}, 27:105103, 2015.

\bibitem{Hennet05}
L.~Hennet, S.~Krishnan, A.~Bytchkov, T.~Key, D.~Thiaudi\`{e}re, P.~Melin,
  I.~Pozdnyakova, M.~L. Saboungi, and D.~L. Price.
\newblock {\em Int. J. Thermophys}, 26:1127, 2005.

\bibitem{Hennet07b}
L.~Hennet, I.~Pozdnyakova, A.~Bytchkov, D.~L. Price, G.~N. Greaves, M.~Wilding,
  S.~Fearn, C.~M. Martin, D.~Thiaudi\`{e}re, J.~F. B\'{e}rar, N.~Boudet, and
  M.~L. Saboungi.
\newblock {\em J. Chem. Phys.}, 126:074906, 2007.

\bibitem{Hennet08}
L.~Hennet, I.~Pozdnyakova, A.~Bytchkov, V.~Cristiglio, D.~Zanghi, S.~Brassamin,
  J.~F. Brun, M.~Leydier, and D.~L. Price.
\newblock {\em J. Non-Cryst. Solids}, 354:5104, 2008.

\bibitem{Bytchkov10}
A.~Bytchkov, L.~Hennet, I.~Pozdnyakova, J.~Wright, G.~Vaughan, S.~Rossano,
  K.~Madjer, and D.~L. Price.
\newblock {\em AIP Conf. Proc.}, 1234:219, 2010.

\bibitem{Hennet11b}
L.~Hennet, I.~Pozdnyakova, A.~Bytchkov, J.~W.~E. Drewitt, J.~Kozaily,
  M.~Leydier, S.~Brassamin, D.~Zanghi, H.~E. Fischer, G.~N. Greaves, and D.~L.
  Price.
\newblock {\em High Temp.-High Press.}, 40:263, 2011.

\bibitem{Skinner13b}
L.~B. Skinner, C.~J. Benmore, J.~K.~R. Weber, M.~C. Wilding, S.~K. Tumber, and
  J.~B. Parise.
\newblock {\em Phys. Chem. Chem. Phys}, 15:8566, 2013.

\bibitem{Labiche07}
J.~C. Labiche, O.~Mathon, S.~Pascarelli, M.~A. Newton, G.~G. Ferre, C.~Curfs,
  G.~Vaughan, A.~Homs, and D.~F. Carreiras.
\newblock {\em Rev. Sci. Instrum.}, 78:091301, 2007.

\bibitem{Neuville19}
D.~R. Neuville.
\newblock personal communication.

\bibitem{Faber65}
T.~E. Faber and J.~M. Ziman.
\newblock {\em Philos. Mag.}, 11:153, 1965.

\bibitem{Drewitt13}
J.~W.~E. Drewitt, C.~Sanloup, A.~Bytchkov, S.~Brassamin, and L.~Hennet.
\newblock {\em Phys. Rev. B}, 87:224201, 2013.

\bibitem{Price89}
D.~L. Price, S.~C. Moss, M.~L. Reijers, R.~Saboungi, and S.~Susman.
\newblock {\em J. Phys.: Condens. Matter}, 1:1005, 1989.

\bibitem{Elliot91}
S.~R. Elliot.
\newblock {\em Nature}, 354:445.

\bibitem{Petkov98}
V.~Petkov, Th. Gerber, and B.~Himmel.
\newblock {\em Phys. Rev. B}, 58:11982, 1998.

\bibitem{Zeidler09}
A.~Zeidler, J.~W.~E. Drewitt, P.~S. Salmon, A.~C. Barnes, W.~A. Crichton,
  S.~Klotz, H.~E. Fischer, C.~J. Benmore, S.~Ramos, and A.~C. Hannon.
\newblock {\em J. Phys.: Condens. Matter}, 21:474217, 2009.

\bibitem{Hennet06}
L.~Hennet, I.~Pozdnyakova, A.~Bytchkov, V.~Cristiglio, P.~Palleau, et~al.
\newblock {\em Rev. Sci. Instrum.}, 77:053903, 2006.

\bibitem{Fischer02}
H.~E. Fischer, G.~J. Cuello, P.~Palleau, D.~Feltin, A.~C. Barnes, Y.~S. Badyal,
  and J.~M. Simonson.
\newblock {\em Appl. Phys. A}, 74:S160, 2002.

\bibitem{Mei08a}
Q.~Mei, C.~J. Benmore, J.~Siewenie, J.~K.~R. Weber, and M.~Wilding.
\newblock {\em J. Phys.: Condens. Matter}, 20:245106, 2008.

\bibitem{Cote92}
B.~Cot\'{e}, D.~Massiot, F.~Taulelle, and J.-P. Coutures.
\newblock {\em Chem. Geol.}, 96:367, 1992.

\bibitem{Florian18}
P.~Florian, A.~Novikov, J.~W.~E. Drewitt, L.~Hennet, V.~Sarou-Kanian,
  D.~Massiot, H.~E. Fischer, and D.~R. Neuville.
\newblock {\em Phys. Chem. Chem Phys.}, 20:27865, 2018.

\bibitem{Dupree97}
R.~Dupree, A.~P. Howes, and S.~C. Kohn.
\newblock {\em Chem. Phys. Lett.}, 276:399, 1997.

\bibitem{Eckersley88}
M.~C. Eckersley, P.~H. Gaskell, A.~C. Barnes, and P.~Chieux.
\newblock {\em Nature}, 335:525, 1988.

\bibitem{Gaskell91}
P.~H. Gaskell, M.~C. Eckersley, A.~C. Barnes, and P.~Chieux.
\newblock {\em Nature}, 350:675, 1991.

\bibitem{Skinner12}
L.~B. Skinner, C.~J. Benmore, J.~K.~R. Weber, S.~Tumber, L.~Lazareva,
  J.~Neuefeind, L.~Santodonato, J.~Du, and J.~B. Parise.
\newblock {\em J. Phys. Chem. B}, 116:13439, 2012.

\bibitem{Sears92}
V.~F. Sears.
\newblock {\em Neutron News}, 3:26, 1992.

\bibitem{Fischer06}
H.~E. Fischer, A.~C. Barnes, and P.~S. Salmon.
\newblock {\em Rep. Prog. Phys.}, 69:233, 2006.

\bibitem{Lacy63}
E.~D. Lacy.
\newblock {\em Phys. Chem. Glasses}, 4:234, 1963.

\bibitem{Iuga05}
D.~Iuga, C.~Morais, Z.~Gan, D.~R. Neuville, L.~Cormier, and D.~Massiot.
\newblock {\em J. Am. Chem. Soc.}, 127:11540, 2005.

\bibitem{Skinner12b}
L.~B. Skinner, A.~C. Barnes, P.~S. Salmon, H.~E. Fischer, J.~W.~E. Drewitt, and
  V.~Honkim\"{a}ki.
\newblock {\em Phys. Rev. B}, 85:064201, 2012.

\bibitem{Neuville08}
D.~R. Neuville, L.~Cormier, D.~De~Ligny, J.~Roux, A.~M. Flank, and P.~Lagarde.
\newblock {\em Am. Mineral.}, 93:228, 2008.

\bibitem{Neuville10}
D.~R. Neuville, G.~S. Henderson, L.~Cormier, and D.~Massiot.
\newblock {\em Am. Mineral.}, 95:1580, 2010.

\bibitem{Turci17}
F.~Turci, C.~P. Royall, and T.~Speck.
\newblock {\em Phys. Rev. X}, 7:031028, 2017.

\end{thebibliography}
\bibliographystyle{unsrt}

\end{document}